\documentclass{article}

\usepackage{amsmath, amsthm, amssymb, amsfonts}
\usepackage{subcaption}
\usepackage[a4paper, total={6.5in, 9.5in}]{geometry}
\input{tikzit.sty}

\usepackage{wrapfig}

\newcommand*{\ket}[1]{\left|#1\right\rangle}

\usepackage{quantikz}
\usetikzlibrary{external}
\usepackage[noblocks]{authblk}
\usepackage{graphicx}
\usepackage{wrapfig}
\usepackage{float}

\usepackage{hyperref}
\usepackage[ruled,vlined]{algorithm2e}
\usepackage[nopatch]{microtype}
\usepackage{booktabs}
\usepackage{stfloats}
\usepackage{caption}
\usepackage[normalem]{ulem}
\usepackage[mode=build]{standalone}
\usepackage{pdflscape}
\usepackage{bm}
\usepackage{xcolor}
\usepackage{listings}

\usepackage{verbatim}
\usepackage{etoolbox}
\usepackage{placeins}

\newtheorem*{lemma*}{Lemma}

\definecolor{keywordcolor}{RGB}{0,0,255}
\definecolor{commentcolor}{RGB}{0,128,0}
\definecolor{stringcolor}{RGB}{163,21,21}

\usepackage[
backend=biber,
maxbibnames=99,
style=alphabetic,url=false,doi=false
]{biblatex}
\lstdefinestyle{python}{
    basicstyle=\ttfamily,
    keywordstyle=\color{keywordcolor},
    stringstyle=\color{stringcolor},
    showstringspaces=false,
    breaklines=true,
    tabsize=4,
    frame=single,
    mathescape=true,
    keywords={assert, for, is, not, Quint, Mod, LookupTable, def, free, int, range, return},
    moredelim=[is][\color{brown}]{"""}{"""},
    commentstyle=\color{commentcolor}\emph
}

\renewcommand{\mod}{\mathbin{\;\mathrm{mod}\;}}
\newcommand{\Tof}{\mathsf{Tof}}
\title{Optimized circuits for windowed modular arithmetic with applications to quantum attacks against RSA}

\author[1,2]{Alessandro Luongo}
\author[3]{Varun Narasimhachar}
\author[4,*]{Adithya Sireesh}

\affil[1]{\small{Centre for Quantum Technologies, National University of Singapore, Singapore}}
\affil[2]{Inveriant Pte. Ltd., Singapore}
\affil[3]{Institute of High Performance Computing, Agency for Science, Technology and Research, Singapore.}
\affil[4]{School of Informatics, University of Edinburgh, Scotland, United Kingdom}
\affil[*]{Corresponding author: \url{asireesh@ed.ac.uk}}

\begin{document}
\maketitle

\begin{abstract}
Windowed arithmetic [Gidney, 2019] is a technique for reducing the cost of quantum arithmetic circuits with space--time tradeoffs using memory queries to precomputed tables. It can reduce the asymptotic cost of modular exponentiation from $\mathcal{O}\left(n^3\right)$ to $\mathcal{O}\left(n^3/\log^2 n\right)$ operations, resulting in the current state-of-the-art compilations of quantum attacks against modern cryptography. In this work we introduce four optimizations to windowed modular exponentiation. We (1) show how the cost of unlookups can be reduced by $66\%$ asymptotically in the number of bits, (2) illustrate how certain addresses can be bypassed, reducing both circuit depth and the overall lookup cost, (3) demonstrate that multiple lookup--addition operations can be merged into a single, larger lookup at the start of the modular exponentiation circuit, and (4) reduce the depth of the unary conversion for unlookups. On a logical level, this leads to a $3\%$ improvement in Toffoli count and Toffoli depth for modular exponentiation circuits relevant to cryptographic applications. This translates to some improvements on [Gidney and Eker\aa, 2021]'s factoring algorithm: for a given number of physical qubits, our improvements show a reduction in the expected runtime from $2\%$ to $6\%$ for factoring $\mathsf{RSA}$-$2048$ integers. 
\end{abstract}

\section{Introduction}

Efficient quantum arithmetic circuits are fundamental to a wide range of quantum algorithms, from cryptographic applications to simulations in physics and chemistry, machine learning, and finance. For example, these circuits feature prominently in Shor's quantum factoring algorithm~\cite{Shor1994}, whose implementation would render many cryptographic schemes, including $\mathsf{RSA}$ (Rivest–Shamir–Adleman) and $\mathsf{ECC}$ (elliptic curve cryptography), insecure. 
Despite the theoretical promise of speedups offered by quantum algorithms, practical implementations demand highly optimized arithmetic circuits to minimize qubit and gate resource overheads. At the core of quantum factoring algorithms is a circuit for the modular exponentiation operation, which dominates the costs\textemdash including physical qubit count, gate count, and runtime\textemdash required for their execution. Hence, optimizing modular arithmetic subroutines within these algorithms is crucial for making quantum attacks on cryptography practically viable. In quantum computing, more significantly than in the classical case, circuit optimizations\textemdash such as reducing the space or the number of gates\textemdash can mean the difference between being able to run an algorithm or not.

In this regard, significant progress has already been made in improving quantum arithmetic algorithms~\cite{draper2000addition, cuccaro2004new, kahanamoku2024fast, litinski2024quantum, gidney2018halving} and their applications, especially in cryptanalysis, e.g., with newer variants of factoring algorithms and more efficient modular arithmetic subroutines to reduce resource overheads~\cite{gidney2021factor, litinski2023compute, gouzien2023performance,chevignard2024reducing,wang2024comprehensive}. There have also been attempts to adapt more efficient classical methods of multiplication (such as the algorithms by Karatsuba, Toom\textendash Cook, and Schönhage\textendash Strassen \cite{knuth2014art}) to the quantum setting. Naively doing so leads to a significant waste of space and time as the quantum counterparts of these subroutines need to be executed reversibly. The asymptotically efficient quantum Karatsuba algorithm~\cite{gidney2019asymptotically} and fast integer multiplication with zero ancillas~\cite{kahanamoku2024fast} are more recent approaches to adapt efficient classical multiplication algorithms to quantum algorithms without the massive space overhead.

One notable advancement in the area of efficient quantum arithmetic is Gidney's introduction of windowed quantum arithmetic~\cite{gidney2019windowed}, which reduces the costs of modular multiplication through precomputed lookup tables and the segmentation of operations into ``windows''. This method improves the scaling of modular exponentiation of $n$-bit numbers from $\mathcal{O}(n^3)$ to $\mathcal{O}(n^3 / \log^2 n)$ by offloading some computational overhead to classical resources. This technique, which is a central focus of optimization in our work, underpins one of the best circuits for factoring $\mathsf{RSA}$-$2048$ integers, demonstrating its utility in practical quantum cryptographic attacks~\cite{gidney2021factor}.

\subsection{Our contribution}
In this paper, we present four improvements for windowed arithmetic circuits for modular multiplication and examine their impact on quantum algorithms for breaking $\mathsf{RSA}$-$2048$. 

In Section~\ref{sec:factoringandquantumarithmetic}, we first look at the history of quantum factoring algorithms, highlighting the key optimizations developed over the past three decades, and their interplay with quantum arithmetic. We discuss the growing need for more efficient and cost-effective subroutines, particularly arithmetic ones, to reduce the overhead of running these algorithms in a fault-tolerant manner. We introduce the background on memory lookups and modular arithmetics in Section~\ref{sec:preliminaries-background}.

In Section~\ref{sec:improvements}, we present four key algorithmic optimizations to the windowed arithmetic method, focusing on reducing the size and depth  (measured in Toffoli gates and Toffoli depth) of memory lookups and their uncomputation, as well as minimizing the number of required lookups in the modular exponentiation algorithm. First, we show how the cost of unlookups can be reduced by $66\%$ asymptotically in the number of bits. Then, we illustrate how certain addresses can be bypassed, reducing both circuit depth and the overall lookup cost. Furthermore, we demonstrate that by merging multiple lookup-addition operations into a single, larger lookup at the start of the modular exponentiation circuit, additional savings in both Toffoli gate count and circuit depth can be achieved. Finally, by using existing techniques, we also show how the depth of unary conversion (a subroutine used in computing memory lookups) can be reduced. While the improvements depend on parameters like the number of bits of the modulus, number of bits of the exponent register, and error budget, in ranges relevant for cryptographic applications, we achieve a $3\%$ improvement in Toffoli count and Toffoli depth, compared to the original windowing arithmetic circuit.

\begin{table}[ht]
\centering
\renewcommand{\arraystretch}{1.5}
\resizebox{\textwidth}{!}{
\begin{tabular}{c||c|c|c|c|c|c|c|c|c}
\hline\hline
                        & $\mathsf{Adt\_factor}$ & $\mathsf{Reps}$ & \multicolumn{3}{c|}{$\#\mathsf{Tof}$} & \multicolumn{3}{c|}{$\mathsf{Depth}$} \\ \hline
                        &      &              & $\mathsf{Lookup}$ & $\mathsf{Add.}$ & $\mathsf{Unlookup}$ & $\mathsf{Lookup}$ & $\mathsf{Add.}$ & $\mathsf{Unlookup}$ \\ \hline
Original~\cite{gidney2019windowed}          &     0 &   $2\frac{nn_e}{w_m w_e}$ & $2^{w_e + w_m}$ & $2n$ & $3\sqrt{2^{w_e + w_m}}$ & $2^{w_e + w_m}$ & $2n$ & $3\sqrt{2^{w_e + w_m}}$  \\ \hline
OPT. 1 [Sec.~\ref{subsec:imp1-deferred}]              &  0   & $2\frac{nn_e}{w_m w_e}$ & $2^{w_e + w_m}$ & $2n$ & $2\frac{w_m}{n}\cdot2^{w_e} + 2^{w_m}$ &  $2^{w_e + w_m}$ & $2n$ & $2\frac{w_m}{n}\cdot2^{w_e} + 2^{w_m}$  \\ \hline
OPT. 2 [Sec.~\ref{subsec:imp3-selective}]                & 0 & $2\frac{nn_e}{w_m w_e}$  & $2^{w_e + w_m}-2^{w_e}$ & $2n$ & $3\sqrt{2^{w_e + w_m}}$ & $2^{w_e + w_m}-2^{w_e}$ & $2n$ & $3\sqrt{2^{w_e + w_m}}$  \\ \hline
OPT. 3 [Sec.~\ref{subsec:imp4-larger}]                 &  $2^{n'_e}$  & $2\frac{n(n_e-n'_e)}{w_m w_e}$& $2^{w_e + w_m}$ &$2n$  &$3\sqrt{2^{w_e + w_m}}$  &  $2^{w_e + w_m}$ &$2n$  &$3\sqrt{2^{w_e + w_m}}$ \\ \hline
OPT. 4 [Sec.~\ref{subsec:imp5-lower-depth-unary}]                  & 0  & $2\frac{nn_e}{w_m w_e}$& $2^{w_e + w_m}$ & $2n$ & $3\sqrt{2^{w_e + w_m}}$ &  $2^{w_e + w_m}$ & $2n$ & $\sqrt{2^{w_e + w_m}} + 2(w_e - 1)$ \\ \hline\hline
OPT. 1+2+3+4          &   $2^{n'_e}$   &  $2\frac{n(n_e-n'_e)}{w_m w_e}$ & $2^{w_e + w_m}-2^{w_e}$ & $2n$  & $2\frac{w_m}{n}\cdot2^{w_e} + 2^{w_m}$ &  $2^{w_e + w_m}-2^{w_e}$ & $2n$  & $2\frac{w_m}{n}(w_e-1) + 2^{w_m}$ \\ \hline\hline
\end{tabular}
}
\vspace{0.2cm}
\caption{Comparison of the computational costs of the original windowed modular exponentiation and each of our proposed optimizations. Here, \( n \) denotes the number of bits in the modulus, \( n_e \) represents the number of exponent bits, and \( w_e \) and \( w_m \) correspond to the sizes of the exponent and multiplication windows, respectively. For OPT 3, \( n'_e \) indicates the number of exponent bits directly exponentiated at the start of the modular exponentiation procedure due to a larger initial lookup. For a specific \( n \), optimal values for \( w_e \), \( w_m \), and \( n'_e \) can be determined via a grid search. The total $\Tof$ count and $\mathsf{depth}$ can be calculated from the table as $\mathsf{Adt\_factor}+ \mathsf{Reps}(\mathsf{Lookup} + \mathsf{Add.}+\mathsf{Unlookup})$. An improvement in number of logical qubits is reported in Appendix~\ref{subsec:imp2-slicedwindowing}}
\label{tab:resource_overheads_all_improvements}
\end{table}

In Section~\ref{sec:application-to-break-rsa}, we study how our improvements translate to saving in physical resources. In particular, we study the improved physical qubit count, runtime, and error-correction overhead, providing updated resource estimates for attacking $\mathsf{RSA}$-$2048$. 
Concretely, we test our improvements within the Gidney\textendash Eker\aa\ (GE) framework~\cite{gidney2021factor} and demonstrate a reduction in the computational volume for factoring $\mathsf{RSA}$-$2048$ integers. Combined, these optimizations yield reductions in the Toffoli count for attacks against $\mathsf{RSA}$ by $1.5\%$ to $3.4\%$, depending on the key size. For a fixed physical qubit count, our improvements show anywhere from a $2\%$ to $6\%$ reduction in the expected runtime for factoring $\mathsf{RSA}$-$2048$ integers (Table~\ref{tab:impact_qecc_superconducting_increasing_qubit_count}). These improvements help find parameters for the original GE algorithm that lead to a slight reduction in the overall computational volume for factoring $\mathsf{RSA}$-$2048$. Finally, we explore potential tradeoffs between runtime and space when integrating our improvements into the GE algorithm. Depending on the chosen cost metric, the qubit count can be reduced by nearly $25\%$ at the expense of a $1.5$ times increase in the runtime. These tradeoffs were already part of the optimization landscape considered in \cite{gidney2021factor}; but even accounting for them, our contributions lead to slight reductions in the associated costs. The tradeoffs are explored in Section~\ref{sec:application-to-break-rsa} (Figure~\ref{fig:all-surviving-estimates}) and Appendix~\ref{apx:mqb_tradeoffs_ge_v_ours}.

In Section~\ref{sec:discussion}, we discuss further ideas, exploring other potential space-time tradeoffs in circuit design. Specifically, we discuss modifications to the memory lookup architecture used in~\cite{gidney2021factor} for a memory with reduced depth at the expense of increased space requirements. In Appendix~\ref{subsec:imp2-slicedwindowing}, we discuss an optimization that leads to a reduction in the number of \emph{logical} qubits used in the windowing operation. However, we were unable to translate these reductions into a corresponding decrease in physical qubits or total runtime. Last but not least, the reader interested in a more precise understanding of the subroutines used in GE factoring algorithm is referred to Appendix~\ref{apx:pseudocode}, where they can find the pseudocode of most of the subroutines discussed in this work.

\subsection{Factoring and quantum arithmetic}\label{sec:factoringandquantumarithmetic}

Shor's algorithm provides an efficient quantum approach to factoring an integer $N$. This problem can be reduced to finding the order $r$ of an integer $b$ in the multiplicative group of integers modulo $N$, where $b < N$ and $\gcd(b, N) = 1$. The order $r$ satisfies the condition
\begin{align*}
    b^r \equiv 1 \mod N.
\end{align*}
If $r$ is even, this condition can be rewritten as
\begin{align*}
    (b^{r/2} - 1)(b^{r/2} + 1) \equiv 0 \mod N.
\end{align*}
Provided $b^{r/2}$ is not a trivial root of unity (i.e., $b^{r/2} \not\equiv \pm1 \mod N$), the factors of $N$ can be extracted using 
\begin{align*}
    \gcd(N, b^{r/2} - 1) \quad \text{or} \quad \gcd(N, b^{r/2} + 1).
\end{align*}

\noindent Shor's quantum factoring algorithm finds $r$ by leveraging phase estimation: a ubiquitous quantum computing subroutine that, when given a prepared eigenstate of a unitary operator, determines the corresponding eigenvalue to a required precision. In this case, the operator is the modular multiplication unitary $\mathsf{U}_b$, which acts as follows on the computational basis:
\begin{align*}
    \mathsf{U}_b \ket{y} = \ket{b \cdot y \mod N}.
\end{align*}
Here, the integers $y$ are represented as binary strings encoded in quantum registers. The computational basis $\ket{y}_{n}$ represents integers $x \in \{0, 1, \dots, 2^{n}-1\}$, where $n$ is the bit size of the number we wish to factorize. The quantum state prepared by the algorithm is of the form: 
\begin{align*}
    \sum_{x=0}^{2^{2n}-1}\ket{x}_{2n}\ket{b^x \mod N}_n.
\end{align*}

\noindent
This state is generated through a sequence of modular multiplication operations, where a chosen integer $b$ (with $\gcd(b, N) = 1$) is used. A QFT (quantum Fourier transform) is then applied to the first register to perform phase estimation, resulting in an approximation of $s/r$, where $0 \leq s < r$. The continued fractions algorithm is subsequently used to extract $r$ from this approximation. For successful extraction, the approximation of $s/r$ must have sufficient precision: approximately $2n$ bits or an error smaller than $1/N^2$. This requirement determines the size of the $x$ register to be $2n$ qubits, ensuring $s/r$ is a convergent of the approximation~\cite{nielsen}.

\paragraph{Variants of Shor's quantum factoring algorithm.}\label{sec:variants-of-quantum-factoring-alg} Beyond efforts to optimize modular multiplication subroutines, several alternative approaches to factoring numbers have been proposed, diverging from Shor's original framework. Numerous studies~\cite{gidney2021factor, litinski2023compute, gouzien2023performance, may2019quantum, ekeraa2017quantum, rines2018high} explore these variations. Eker\aa\ and H{\aa}stad for instance, found that it is easier to factor $\mathsf{RSA}$ integers (integers of the form $N=pq$ for large primes $p, q$) by reducing the problem of factoring to a short discrete logarithm problem~\cite{ekeraa2017quantum}. May and Schlieper~\cite{may2019quantum} proved that Shor's algorithm is compression-robust and that the target state $\ket{b^x}$ can be hashed to a single bit (at the cost of more repetitions of the algorithm). This result was then extended by \textcite{chevignard2024reducing} by using a residue number system modular multiplier~\cite{rines2018high} to reduce the space requirements of factoring an $n$-bit integer to just $n/2 + o(n)$. In 2024, Regev \cite{regev2023efficient} combined different lattice-based techniques to come up with a multi-dimensional analogue of Shor's algorithm, requiring $\Tilde{\mathcal{O}}(n^{3/2})$ gate cost and $\Tilde{\mathcal{O}}(n^{3/2})$ space at the cost of $\Tilde{\mathcal{O}}(n^{1/2})$ repetitions. The space requirements were further relaxed by Ragavan and Vaikuntanathan~\cite{ragavan2023optimizing} to $\mathcal{O}(n)$ using a Fibonacci exponentiation technique (an optimization based purely on making multiplication easier in Regev's algorithm), at the cost of increasing the gate cost to $\mathcal{O}(n^{5/2})$. Eker\aa\ and Gartner~\cite{ekeraa2024extending} then extended Regev's work to computing discrete logarithms. They also showed a way to use Regev's algorithm to solve the order-finding problem. A very early work that inspired the idea of trading more repetitions for reduced space (as seen in many algorithms described above) is that of Seifert~\cite{seifert2001using}, which managed to lower the size of the exponent from $2n$ to $n(1+\epsilon)$ for $0<\epsilon\leq1$ and computing approximations to the order of $b$. These approximations are then combined using simultaneous diophantine approximations to reconstruct the order. While the factoring algorithms by Regev and \textcite{chevignard2024reducing} offer interesting avenues to explore the practical costs of breaking $\mathsf{RSA}$, there is still work to be done in further reducing space requirements in Regev's algorithm and gate/time costs in \textcite{chevignard2024reducing}'s approach. In this paper, we focus on the implementation of factoring $\mathsf{RSA}$ integers using the GE algorithm and explore the impact of various improvements to the current state of the art.

\paragraph{Gidney\textendash Eker\aa\ (GE) 2021.}
This relatively recent work is among the more thorough analyses of concrete resources (qubits, gates, time, etc.)\ required for running a quantum algorithm on specific, practically relevant instances of a problem (as opposed to asymptotic analyses). The work lays out a circuit compilation of Shor's factoring algorithm for attacking state-of-the-art $\mathsf{RSA}$ cryptographic schemes. The compilation is tailored for superconducting qubit architectures with a layout suitable for error correction using surface codes. For relevant $\mathsf{RSA}$ key sizes, the work optimizes various design parameters of the circuit (which we will explain in more detail below) and estimates the corresponding resource costs based on realistic assumptions on near-term hardware. In addition to its detailed resource estimation, it also presents some improvements in some subroutines that occur in the algorithm. A key focus of their approach—like many algorithms optimizing fault-tolerant quantum computation—is the reduction of $\Tof/T$ gate count and depth, as these gates dominate the cost of implementing large-scale quantum algorithms~\cite{fowler2018low,gidney2019flexible, babbush2018encoding,Haner2020}. In this section, we detail the main algorithmic techniques used by Gidney and Eker\aa ~\cite{gidney2021factor} to achieve a physical qubit count of $20$ million qubits and expected runtime of $8$ hours for breaking $\mathsf{RSA}$-$2048$ keys:
\begin{itemize}
    \item \emph{Factoring using discrete logarithm problem}: Eker\aa\ and H{\aa}stad showed that $\mathsf{RSA}$ integers could be factored more efficiently than Shor's algorithm by translating the factoring problem to a short discrete logarithm problem \cite{ekeraa2017quantum}. The reduction of the $n$-bit factoring problem to a discrete logarithm problem helps reduce the number of exponentiation qubits to $1.5n$ (we normally require at least $2n$~\cite{Shor1994}).
    \item \emph{Windowed arithmetic via lookups}: Previously, it was thought that each controlled modular multiplication in Shor's algorithm would have to be performed individually. However, in the paper ~\cite{gidney2019windowed}, it was shown that we can precompute a table of values classically and appropriately load them into our circuit. This method leads to a $\log^2 n$ factor reduction in the number of Toffolis required for modular exponentiation from $\mathcal{O}(n^3)$ to $\mathcal{O}(n^3/\log^2 n)$.
    \item \emph{Semi-classical Fourier transform}: The QFT that needs to be performed at the end of the phase estimation in Shor's algorithm  involves $\mathcal{O}(n^2)$ controlled rotation gates. There is, however, a way to perform the same procedure semi-classically ~\cite{griffiths1996semiclassical}: the exponent qubits can be measured one at a time, and each measurement outcome can be used to classically control a rotation on the subsequent exponent qubits. This technique also allows for reuseing the exponent qubits instead of maintaining a superposition of $\mathcal{O}(n)$ qubits at one time.
    \item \emph{Coset representation}: A basic modular adder requires the use of 4 non-modular adders. Adding two $n$-qubit registers has a Toffoli cost of $2n$. When we need to perform modular addition, the cost goes up to $8n$. The coset representation, first introduced by Zalka \cite{zalka2006shor}, helps perform modular addition with a single non-modular addition circuit. For a register $\ket{x}$ that we wish to add into, we first map it into the state $\sum_c^{\mathcal{O}(\log N)} \ket{x + c*N}$ (the coset state).  This state is close to the eigenvector of the unitary ``add $N$ to $x$''. Now, if we wish to add a value $y$ into $\ket{x'}$, we just perform non-modular addition into $\sum_{c=1}^{2^k} \ket{x + c\cdot N + y}_{n+k} \approx \sum_{c=1}^{2^k} \ket{(x+y\;\mod N)  +  c\cdot N}$. The number of extra qubits required, $k$, is logarithmic in the total number of modular additions to the performed in the whole algorithm.
    \item \emph{Oblivious carry runways}: The depth of the ripple carry adder is limited by the size of the registers that are to be summed because the more significant qubits need to wait for the carry values to be propagated from the less significant qubits. Carry runways~\cite{gidney2019approximate} give us a way to parallelize addition by splitting the addition registers into multiple pieces. The pieces of the first register are summed up with the corresponding pieces of the second register. Thus, the time taken by addition now effectively depends only on the size of the pieces (instead of the size of the whole register).
\end{itemize}

\section{Preliminaries and background}\label{sec:preliminaries-background}
We denote quantum registers as collections of qubits used for storing quantum information. For example \( \ket{x}_n \) refers to an \( n \)-qubit register in the computational basis with value \( x \in \{0, 1, \dots, 2^n - 1\} \). We use $x_i$ to denote the $i^{\text{th}}$ qubit of the register $x$. In windowed arithmetic, we split a register \( \ket{x}_n \) into a consecutive set of qubits or ``windows'' of size \( w \) qubits, with each window denoted as \( x_{(i,w)} \) for \( i = 0, 1, \dots, n / w - 1 \). Note: if $w$ does not divide $n$ the last window size has $n \mod w$ bits. However, for ease of explanation, we assume that $w$ divides $n$. Each \( x_{(i,w)} \) is treated as an independent \( w \)-bit value. For instance, the overall value of
$x$ can be expressed in terms of $w$-bit windows as follows:
\[
x = \sum_{i=0}^{n/w - 1} x_{(i,w)} \cdot 2^{i \cdot w}.
\]
Using this notation to represent windows, we see that $x_{(i,1)}\equiv x_i$ i.e.\ the $i^{\text{th}}$ qubit of register $x$ is equivalent to saying that we are looking at the $i^{\text{th}}$ window of register $x$ with window size 1.

\begin{wrapfigure}{r}{0.45\textwidth}
\centering
    \centering
    \includegraphics[width=0.35\textwidth]{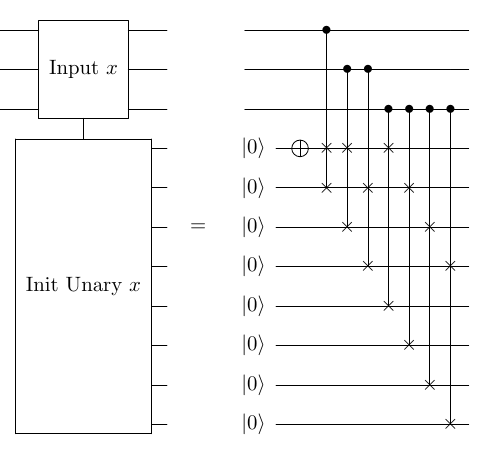}
    \caption{Unary conversion from~\cite{gidney2019windowed}. This circuit essentially performs a binary to unary encoding by controlling on the least significant qubit and shifting the position by 1 in the unary register. The next pair of gates control on the 2\textsuperscript{nd} least significant qubit, shifting the position by 2 places. The last set of 4 controlled swaps control on the most significant qubit and shift by 4 positions.}
    \label{fig:unary_conversion}
\end{wrapfigure}

\paragraph{Unary encoding.}
We define a unary encoding  (also referred to as one-hot encoding in other fields, such as machine learning) as a function \(\mathsf{unary}: \{0,1\}^n \to \{0,1\}^{2^n},\) where an \( n \)-bit input \( x \) is mapped into a \( 2^n \)-bit output. The mapping is defined as $\mathsf{unary}(x) = \mathsf{bin}(2^{x+1})$, where $\mathsf{bin}$ is the function returning the binary encoding of a decimal number with the proper padding. 
In this encoding, the position of the bits set to $1$ corresponds directly to the value of the binary input \( x \). The possible circuit is illustrated in Figure~\ref{fig:unary_conversion}, involving controlled operations that shift the position of the $1$ in the unary register. The unary register is initialized with the first qubit in the \( 1 \) state and all other qubits in the \( 0 \) state. This configuration effectively represents the value '0' in the unary register. The least significant bit of the input register \( x \) controls the first shift, moving the $1$ by one position if the bit is set. Subsequent gates control on the higher-order bits of \( x \), where each \( i^{th} \) bit shifts the position of the $1$ by \( 2^{i-1} \) positions when set. This approach encodes an \( n \)-bit binary input into a unary register of size \( 2^n \), positioning the $1$ according to the binary value of \( x \). The depth and $\mathsf{Tof}$ count of this approach is $2^n-1$.

\paragraph{Computational model and circuit complexities.}
We cost our circuits by measuring the size as the number of Toffoli ($\mathsf{Tof}$) gates, and the circuit depth as the Toffoli $\#\mathsf{depth}$ (the number of sequential layers of $\mathsf{Tof}$ gates required in the circuit). This is justified as the Toffoli gates are generally considered the main bottlenecks for fault-tolerant quantum computation. Hence, we will use these as a proxy to describe the overall circuit size and depth. In our case, the size and depth of the circuits we consider are similar up to a small multiplicative factor of the Toffoli size and depth. In the windowed arithmetic circuits, we also use Fan-Out gates, which are described as single control multi-target NOT gates. In the standard circuit model, we can decompose a Fan-Out gate of arity $k+1$ using $k$ CNOT gates and depth $O(\log_2(k))$. Another way of implementing Fan-Out gates can be found in~\cite{pham20132d}, with a protocol that requires $O(k)$ qubits and a constant number of operations. In a more realistic, error-corrected setting, Fan-Out gates can be implemented using lattice surgery~\cite{horsman2012surface,fowler2018low} as a single logical operation. In particular, in~\cite{gidney2019flexible}, they explore layouts for surface codes to reduce time overheads for performing large Fan-Out gates (as in the case of QROM lookup table operations). We apply a more realistic method of costing the computation in Section~\ref{sec:application-to-break-rsa} as described in~\cite{gidney2021factor, gidney2019flexible}.

\subsection{Quantum table lookup and unlookup}\label{sec:lookups}

\begin{figure}[!ht]
    \centering
    \includegraphics[width=\textwidth]{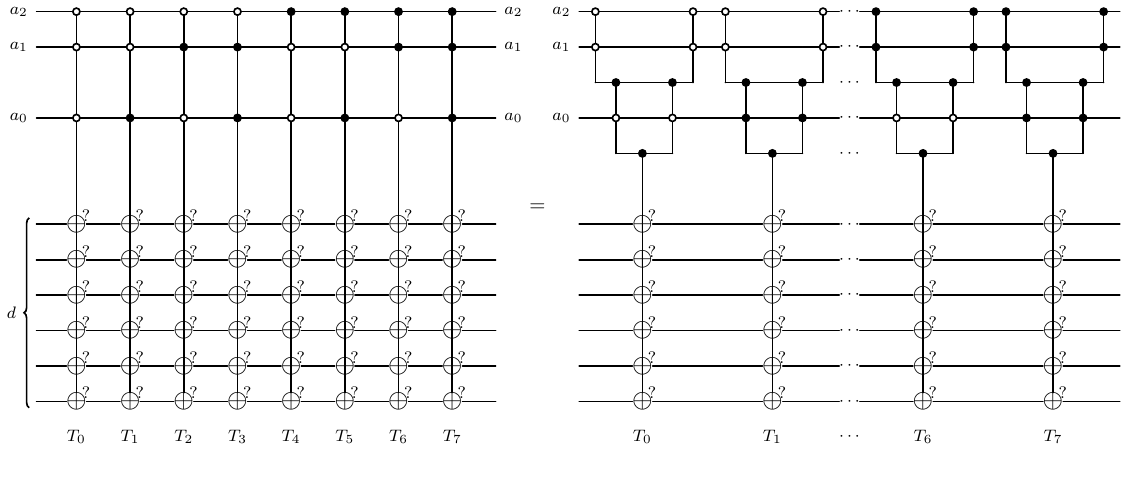}
    \caption{Equivalence of a standard, and unoptimized QROM lookup, and the same table decomposed into a set of logical-$\mathsf{AND}$'s. The two-control gates that look like $\mathsf{Tof}$ gates can be decomposed in the form as shown in Figs.~\ref{fig:gidney-temp-logical-and}, ~\ref{fig:gidney-temp-logical-and-uncomputation}}
    \label{fig:standard-decomposed-qrom-lookup}
\end{figure}

While the focus of this paper is on optimizing windowed arithmetic, we begin by introducing quantum table lookups—a key operation within windowed arithmetic. These lookups enable the efficient loading of classical or quantum data into memory~\cite{babbush2018encoding}. In this section, we provide a detailed exploration of table lookups as a foundation for the subsequent discussion on windowed arithmetic. For an address register $\ket{a}_l$, we have an associated lookup table of size $L=2^l$. Let us assume that each memory element in the lookup table is of $m$ bits; therefore, we need $m$ qubits to store the value being looked up. A lookup operation can be described as follows:

\begin{align*}
\ket{a}_{l}\ket{y}_m\xmapsto{\mathsf{Lookup}}\ket{a}_{l}\ket{y\oplus T_a}_m,
\end{align*}

\noindent
where $T_a$ is the memory element in the lookup table for index $a$. A specific architecture for a quantum lookup table, termed a $\mathsf{QROM}$ lookup, was introduced in~\cite{babbush2018encoding}; an unoptimized version of the $\mathsf{QROM}$ lookup can be seen in Figure~\ref{fig:standard-decomposed-qrom-lookup}, left.

\begin{figure}[!ht]
    \centering
    \includegraphics{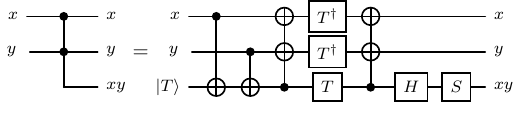}
    \caption{
        The temp logical-$\mathsf{AND}$ construction of Gidney~\cite{gidney2018halving}. Here, the decomposition is using 4 $\mathsf{T}$ gates instead of the more expensive 7 $\mathsf{T}$ gates traditionally used to execute a $\mathsf{Tof}$ gate}
    \label{fig:gidney-temp-logical-and}
\end{figure}

Decomposing each multi-controlled Toffoli gate, 
and using a temporary logical-$\mathsf{AND}$ construction (see Figure~\ref{fig:gidney-temp-logical-and}), we get the circuit in Figure~\ref{fig:standard-decomposed-qrom-lookup}, right. This circuit can be further simplified as showing in ~\cite{babbush2018encoding}, resulting in $\Tof$ gate count and $\mathsf{depth}$ of $2^l - 1$.

\begin{figure}[!ht]
    \centering
    \includegraphics{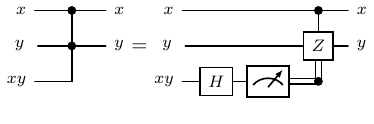}
    \caption{
        Uncomputation of the temp logical-$\mathsf{AND}$ by Gidney~\cite{gidney2018halving}. Here the circuit uses no $\mathsf{T}$ gates. It instead includes an $\mathsf{H}$ gate, a computation basis measurement and a classically controlled $\mathsf{C\text{-}Z}$ gate}
    \label{fig:gidney-temp-logical-and-uncomputation}
\end{figure}

\paragraph{Uncomputation: motivation and improvements.}
In quantum algorithms and subroutines, uncomputation is sometimes necessary, especially when ancillary qubits need to be reset to free up space for reuse. Bennett's method~\cite{Bennett1973} offers a foundational approach to uncomputation, though it may double the resource cost relative to the original computation. Over time, various approaches, both general~\cite{jones2013low, kornerup2021tight} and specific~\cite{babbush2018encoding,gidney2019approximate,gidney2019windowed,gidney2018halving,luongo2024measurement}, have aimed to reduce this overhead.

In the case of quantum table lookups, once a lookup register has been consumed by a subroutine in our algorithm, uncomputation or an unlookup becomes essential, as resetting the corresponding registers saves valuable space. As with other uncomputations, under certain conditions, it admits a simpler implementation than the lookup itself. We now discuss a particular simplification due to Gidney~\cite{gidney2019windowed}, built on the more common MBU technique. The lookup register is first measured in the $\mathsf{X}$ basis, and the resultant measurement value $s$ is recorded
\begin{align*}
\sum_{a}\ket{a}_{\ell}\ket{T_a}_m\xmapsto{\mathsf{Measure}}&\sum_{a}(-1)^{s\cdot T_a}\ket{a}_{\ell}\ket{0}_m.
    \end{align*}

This measurement may result in a phase on each address $a$, which depends on the value of the measurement $s$ and the value of the memory element $T_a$ associated with $a$. These values are all known classically, as we have access to the entire lookup table. We now split the address register in two pieces $a_\mathsf{low}, a_\mathsf{high}$ of $u$ and $\ell - u$ qubits respectively (for $0 < u < \ell$), and use $a_\mathsf{low}$ as the input for the unary conversion.
\begin{align*}
    \sum_{a}(-1)^{s\cdot T_a}\ket{a}_{\ell}\ket{0}_m\xmapsto{\mathsf{Unary}}&\sum_{a}(-1)^{s\cdot T_a}\ket{a=a_{\mathsf{high}}a_{\mathsf{low}}}_{\ell}\ket{\mathsf{unary}(a_\mathsf{low})}_{2^u}\ket{0}_{m-2^{u}}\\
    =&\sum_{a_\mathsf{high}}\ket{a_\mathsf{high}}_{\ell-u}\left[\sum_{a_\mathsf{low}}(-1)^{s\cdot T_{a_\mathsf{high}a_\mathsf{low}}}\ket{a_\mathsf{low}}_{u}\ket{\mathsf{unary}(a_\mathsf{low})}_{2^u}\right]\ket{0}_{m-2^{u}}.
    \end{align*}
The unary conversion is used to construct a new phase correction lookup table $F_{a_{\mathsf{high}}}$, with the address register $a_\mathsf{high}$, and lookup register storing the unary conversion. For an index $a_{\mathsf{high}}$, the value has a $1$ at index $x$, for all $a=a_\mathsf{high}x$ that satisfy $s\cdot T_a=1$. With this new lookup table $F$, we can perform a simultaneous phase correction for any addresses $a=a_\mathsf{high}...$ with the same $a_\mathsf{high}$. The cost of performing the lookup $F$ is $2^{\ell - u}$. We finally perform an inverse unary operation, to reset the unary register, and retrieve the required state.
    \begin{align*}\xmapsto{\mathsf{Lookup}}&\sum_{a_\mathsf{high}}\ket{a_\mathsf{high}}_{\ell-u}\left[\sum_{a_\mathsf{low}}\ket{a_\mathsf{low}}_{u}\ket{\mathsf{unary}(a_\mathsf{low})}_{2^u}\right]\ket{0}_{m-2^{u}} \\ 
    \xmapsto{\mathsf{Unary}^\dagger}&\sum_a \ket{a}_\ell \ket{0}_m.
\end{align*}

The total cost of this unlookup is equal to the sum of the cost of the unary conversion, the cost of the phase correction lookup, and the cost of the unlookup. The unlookup can be done with a temporary logical-$\mathsf{AND}$ construction, thus leading to a total cost of $2^u + 2^{\ell - u}$ $\mathsf{Tof}$ gates. This value is minimized when $u=\ell/2$ i.e.\ when $\mathsf{size}(a_\mathsf{high})=\mathsf{size}(a_\mathsf{low})$, leading to a cost, $2\cdot 2^{\ell/2} = 2\sqrt{L}$ i.e the unlookup has a square root speed up over the cost of the lookup.

\begin{figure}[!ht]
    \centering
    \includegraphics[width=\textwidth]{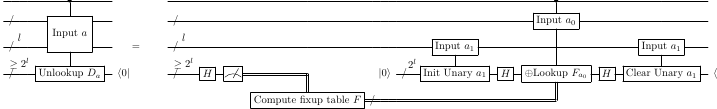}
    \caption{Unlookup circuit from~\cite{gidney2019windowed}.}
    \label{fig:unlookup-circuit}
\end{figure}
\begin{figure}
    \centering
    \includegraphics[width=\textwidth]{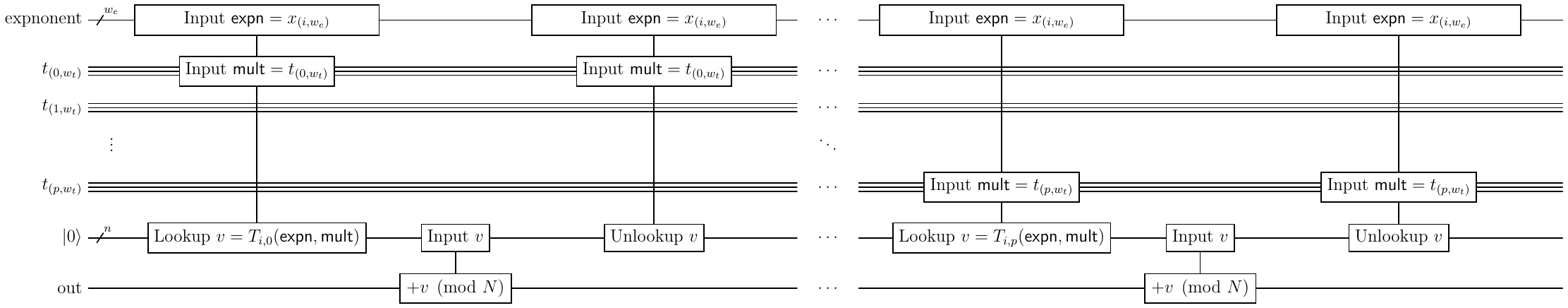}
    \caption{Windowed modular exponentiation by~\cite{gidney2019windowed}. Here, we are executing the modular multiplication of $b^{2^{iw_e}x_{(i,w_e)}}$ (the exponentiated value of the  of the $i^{\text{th}}$ window $x_{(i,w_e)}$ of the exponent), with the register $t=b^{x_{(0,iw_e)}}$ which is currently holding the exponentiation of the first $i-1$ windows of the exponent register.}
    \label{fig:windowing-original}
\end{figure}
In summary, the reduction is based on the above observation that the Toffoli cost of a table lookup is linear in the number of entries but indifferent to the size of each. In the unlookup stage, half of the address register is copied \emph{in unary format} onto part of the lookup (which is now free, thanks to the preceding measurement). Meanwhile, a ``fixup table'' is computed, indexed by the remaining half of the address (therefore, of size quadratically smaller than the original table's) and containing the appropriate phase correction to be applied on each value of the half in unary encoding. This correction is then applied bitwise on the unarized half-address, through a lookup operation over the smaller table. This quadratically reduces the overall Toffoli count of the unlookup.

\subsection{Modular exponentiation using windowed arithmetic}\label{subsec:windowing} 

As mentioned previously, windowed arithmetic reduces the complexity of modular exponentiation from $\mathcal{O}(n^3)$ to $\mathcal{O}(n^3/\log^2n)$. Here we discuss the work of \cite{gidney2019windowed} in detail, which is needed to understand our improvements. We will see why this is the case in the current section. Windowed arithmetic uses lookup tables to perform quantum modular exponentiation of an integer $b$ i.e.\ to prepare the state $\sum_x \ket{x}\ket{b^x \mod N}$.
It is insightful to rewrite the modular exponentiation of $b$ to the power of $x$ by looking at the binary expansion of $x$:
\begin{align*}
    b^x \mod N &= b^{\sum_{i=0}^{k-1} 2^i x_i} \mod N = \prod_{i=0}^{k-1} b^{2^i x_i} \mod N .
    \end{align*}
    
From the above equation, we see that modular exponentiation can be implemented as repeated controlled modular multiplication: namely, multiplying sequentially by $b^{2^i}$ using $x_i$ as the control. We can redefine the computation of the modular exponentiation recursively as follows:
\begin{equation*}
t^{[i]} = t^{[i-1]}b^{2^i x_i} \mod N,
\end{equation*}
with $t^{[0]}=b^{2^0 x_0}$ and $t^{[k-1]}=b^x\mod N$
; Hence, to perform the modular multiplication described previously, we can implement a unitary that takes an integer register $t$ holding the value $t^{[i-1]}$, and computes the value $t^{[i]}=b^{2^ix_i}t^{[i-1]}$ into an empty target register (out-of-place multiplication), for some classically provided $b$ and $i$. More formally, we would like to perform the modular multiplication operation: 

    \begin{align*}
    \ket{x_i}_1 \ket{t^{[i-1]}}_n \ket{c}_n \xmapsto{+b^{2^ix_i}t^{[i-1]}\mod N} & \begin{cases}
        \ket{x_i}_1 \ket{t^{[i-1]}}_n \ket{c + t^{[i]} \mod N}_n & \text{if } x_i = 1 \\
        \ket{x_i}_1 \ket{t^{[i-1]}}_n \ket{c+t^{[i-1]}\mod N}_n & \text{if } x_i = 0,
    \end{cases} \\ 
    \end{align*}
for any $c \in \{0,1\}^n$.     
The text above the arrow depicts the change effected on the last register. For our purposes it suffices to consider $c=0$, thus giving us the operation:
    \begin{align*}
    \ket{x_i}_1 \ket{t^{[i-1]}}_n \ket{0}_n \xmapsto{+b^{2^ix_i}t^{[i-1]}\mod N} & \begin{cases}
        \ket{x_i}_1 \ket{t^{[i-1]}}_n \ket{t^{[i]}}_n & \text{if } x_i = 1 \\
        \ket{x_i}_1 \ket{t^{[i-1]}}_n \ket{t^{[i-1]}}_n & \text{if } x_i = 0.
    \end{cases} \\ 
    \end{align*}

Controlled modular multiplication itself can be decomposed into repeated controlled modular addition, with the controls being $(x_i,t_j)$ for $j\in \{0,1,\cdots,n-1\}$. In words, we iterate over the bits $t_j$ of $t$, for each $j$ adding into the current value of the third register the product of $2^jt_j$ and $b^{2^ix_i}$:
\begin{align*}
\ket{x_i}_1 \ket{t^{[i-1]}}_n \ket{0}_n
         \xmapsto{+ \left(b^{2^i x_i}\right)\cdot t_0 \mod N}& \ket{x_i}_1 \ket{t^{[i-1]}}_n\ket{b^{2^i x_i}t_0\mod N}_n \\
        \xmapsto{+ \left(b^{2^i x_i}\right)\cdot 2t_1 \mod N}& \ket{x_i}_1 \ket{t^{[i-1]}}_n\ket{b^{2^i x_i}(t_0 + 2t_1) \mod N}_n \\
        &\vdots \\
        \xmapsto{+ \left(b^{2^i x_i}\right)\cdot 2^jt_j \mod N}& \ket{x_i}_1 \ket{t^{[i-1]}}_n\ket{b^{2^i x_i}\left(\sum_{k=0}^j 2^k t_k\right) \mod N}_n\\
        &\vdots \\
        \xmapsto{+ \left(b^{2^i x_i}\right)\cdot 2^{n-1}t_{n-1} \mod N}& \ket{x_i}_1 \ket{t^{[i-1]}}_n\ket{b^{2^i x_i}\left(\sum_{k=0}^{n-1} 2^k t_k\right) \mod N}_n\\
        =&\ket{x_i}_1 \ket{t^{[i-1]}}_n\ket{b^{2^i x_i}t^{[i-1]}\mod N}_n.\\
        =&\ket{x_i}_1 \ket{t^{[i-1]}}_n\ket{t^{[i]}}_n.\\
\end{align*}

The acute reader will have observed that every step of the controlled addition can actually be performed by first doing a lookup, then an (uncontrolled) modular addition, which is followed by an unlookup. The lookup table is indexed by $two$-bit addresses $(x_i,t_j)$ and is shown in Table~\ref{tab:lookuptableexample}.  Each operation in the series of lookups followed by additions are  aptly termed lookup-additions or $\mathsf{LookupAdd}$.   

\begin{wraptable}{r}{0.45\textwidth}
\centering
\renewcommand{\arraystretch}{1.5}
\setlength{\tabcolsep}{8pt}
\begin{tabular}{|c|c|c|}
    \hline
    $t_j$ & $x_i$ & $(b^{2^i x_i}) \cdot 2^j t_j \mod N$ \\ \hline
    0 & 0 & 0 \\ \hline
    0 & 1 & 0 \\ \hline
    1 & 0 & $2^j$ \\ \hline
    1 & 1 & $2^j b^{2^i} \mod N$ \\ \hline
\end{tabular}
\caption{A 2-bit lookup table for modular multiplication. Here, $t_j$ is the $j^{\text{th}}$ bit of the multiplicand and $x_i$ is the $i^{\text{th}}$ bit of the exponent.}
\label{tab:lookuptableexample}
\end{wraptable}

Now using lookup tables, we can perform a series of $2$-bit lookup-additions to execute the modular multiplication operation. Since each addition modulo $N$ costs $\mathcal{O}(n)$, and we have to iterate over $n$ bits of $t$, therefore modular multiplication costs $\mathcal{O}(n^2)$. This only exponentiates 1 bit of the exponent and has to be repeated $k$ times for the $k$ different bits of the exponent $x$. However, since all we are doing is performing table lookups followed by additions, we do not need to stop at just looking up 2 bits at a time. Let us assume we take a window $w_e$ of bits for the exponent, and a window $w_t$ of bits for our multiplication register $t$. We can rewrite our recursive definition as:
\begin{equation*}
    t^{[i]} = t^{[i-1]}b^{2^{iw_e} x_{(i,w_e)}} \mod N
\end{equation*}
with $t^{[0]}=b^{2^0 x_{(0,w_e)}}$ and $t^{[k-1]}=b^x\mod N$; where $k=n/w_e$.
We now show how to exponentiate $w_e$ bits at a time. In the next equation \( x_{(i, w_e)} \) represents the \( i \)-th window of size \( w_e \) in the exponent and \( t_{(k, w_t)} \) represents the \( k \)-th window of size \( w_t \) in the multiplication register.

{\small
\begin{align*}
\ket{x_{(i, w_e)}} \ket{t^{[i-1]}}_n \ket{0}_n\xmapsto{+ \left(b^{\left(2^{i \cdot w_e}\right) x_{(i, w_e)}}\right) \cdot t_{(0, w_t)} \mod N} & \ket{x_{(i, w_e)}} \ket{t^{[i-1]}}_n \ket{b^{\left(2^{i \cdot w_e}\right) x_{(i, w_e)}} t_{(0, w_t)} \mod N}_n \\
    \xmapsto{+ \left(b^{\left(2^{i \cdot w_e}\right) x_{(i, w_e)}} \right) \cdot 2^{w_t} t_{(1, w_t)} \mod N} & \ket{x_{(i, w_e)}} \ket{t^{[i-1]}}_n \ket{b^{\left(2^{i \cdot w_e}\right) x_{(i, w_e)}} (t_{(0, w_t)} + 2^{w_t} t_{(1, w_t)}) \mod N}_n \\
    & \vdots \\
    \xmapsto{+ \left(b^{\left(2^{i \cdot w_e}\right) x_{(i, w_e)}} \right) \cdot 2^{j \cdot w_t} t_{(j, w_t)} \mod N} & \ket{x_{(i, w_e)}} \ket{t^{[i-1]}}_n \ket{b^{\left(2^{i \cdot w_e}\right) x_{(i, w_e)}} \left( \sum_{k=0}^{j} 2^{k \cdot w_t} t_{(k, w_t)} \right) \mod N}_n \\
    & \vdots \\
    \xmapsto{+ \left(b^{\left(2^{i \cdot w_e}\right) x_{(i, w_e)}} \right) \cdot 2^{\left(\frac{n}{w_t} - 1\right) \cdot w_t} t_{\left(\frac{n}{w_t} - 1, w_t\right)} \mod N} & \ket{x_{(i, w_e)}} \ket{t^{[i-1]}}_n \ket{b^{\left(2^{i \cdot w_e}\right) x_{(i, w_e)}} \left( \sum_{k=0}^{\frac{n}{w_t} - 1} 2^{k \cdot w_t} t_{(k, w_t)} \right) \mod N}_n \\
    =& \ket{x_{(i, w_e)}} \ket{t^{[i-1]}}_n \ket{b^{\left(2^{i \cdot w_e}\right) x_{(i, w_e)}} t^{[i-1]} \mod N}_n\\
    =& \ket{x_{(i, w_e)}} \ket{t^{[i-1]}}_n \ket{t^{[i]}}_n.
\end{align*}}%

There is one last important ingredient to discuss to finish our presentation of windowed arithmetic. We have discussed how windowing works to perform exponentiation of multiple bits at a time, but we also need to uncompute the register holding $t^{[i-1]}$ (as the multiplication is being performed \emph{out-of-place}), and swap it with the result register, in order to prepare for the next window of exponents to be multiplied. This reset operation is essentially a modular division operation, but can be executed as another modular multiplication operation with the target register being the new multiplication register, and the multiplication register taking the place of the new target register. This works because
\begin{align*}
t^{[i-1]}-\left[b^{-\left(2^{i \cdot w_e}\right) x_{(i, w_e)}}t^{[i]}\right]&=t^{[i-1]}-\left[b^{-\left(2^{i \cdot w_e}\right) x_{(i, w_e)}}b^{\left(2^{i \cdot w_e}\right) x_{(i, w_e)}}t^{[i-1]}\right]\\
&=t^{[i-1]}-t^{[i-1]}=0.    
\end{align*}

The above sequence of computations is as follows:

\begin{align*}\label{eq:muplicand-computation-uncomputation}
\ket{x_{(i,w_e)}}_{w_e}\ket{t^{[i-1]}}_n\ket{0}_{2n}\xmapsto{\mathsf{LookupMul}}& \ket{x_{(i,w_e)}}\ket{t^{[i-1]}}_n\ket{b^{\left(2^{i \cdot w_e}\right) x_{(i, w_e)}}t^{[i-1]}}_n\ket{0}_{n} = 
\ket{x_{(i,w_e)}}\ket{t^{[i-1]}}_n\ket{t^{[i]}}_n\ket{0}_{n}
\nonumber\\
\xmapsto{\mathsf{LookupInvMul}}&\ket{x_{(i,w_e)}}\ket{t^{[i-1]} - t^{[i]}\cdot b^{-\left(2^{i \cdot w_e}\right)x_{(i,w_e)}}}_n\ket{t^{[i]}}_n\ket{0}_{n}\\
    =& \ket{x_{(i,w_e)}}\ket{0}_n\ket{t^{[i]}}_n\ket{0}_{n}.\\
    \end{align*}
We now swap the register holding the result, with the register that was initially holding $t$ ($t$ is now uncomputed because of the $\mathsf{LookupInvMul}$). The result register will become the multiplication register for the next iteration
    \begin{align*}
\xmapsto{\mathsf{Swap}}&\ket{x_{(i,w_e)}}\ket{t^{[i]}}_n\ket{0}_{2n}.
\end{align*} 

Now that the current window ($i^{\text{th}}$ window) has been exponentiated, we can do another round of lookup additions to exponentiate next window i.e.\ the $(i+1)^{\text{th}}$ window of the exponent, until the last remaining exponent window (i.e.\ the $k^{\text{th}}$ exponent window). 

\begin{align*}
\ket{x_{(i+1,w_e)}}_{w_e}\ket{t^{[i]}}_n\ket{0}_{2n} \xmapsto{\mathsf{LookupMul}} & \ket{x_{(i+1,w_e)}}\ket{t^{[i]}}_n\ket{t^{[i+1]}}\ket{0}_{n}\xmapsto{\mathsf{LookupInvMul, Swap}}\ket{x_{(i+1,w_e)}}\ket{t^{[i+1]}}_n\ket{0}_{2n} \\
& \vdots \text{ exponentiation over all $k$ exponent qubit windows} \\
& \vdots \\
\xmapsto{\mathsf{LookupMul, LookupInvMul, Swap}} & \ket{x_{\left(k-1,w_e\right)}}\ket{t^{[k-1]}}\ket{0}_{2n} \\
= & \ket{x} \ket{b^x \mod N}_n\ket{0}_{2n}.
\end{align*}

Thus resulting in the modular exponentiation of $b$. Each step in the repeated controlled modular addition (with controls $x_{(i, w_e)}, t_{(k, w_t)}\; \forall k \in \{0,1,\cdots, n/k - 1\}$) can be performed using a lookup table (as previously shown with the modular addition of $2$ bits) with indices $t_{(k, w_t)}||x_{(i, w_e)}$, where $||$ represents the concatenation of the 2 individual addresses. The size of this lookup table is $L=2^{w_t+w_e}$, and holds memory elements of the form $T_{i,j}(\mathsf{expn},\mathsf{mult})=\left(a^{\left(2^{i \cdot w_e}\right)\mathsf{expn}} \right) \cdot 2^{j \cdot w_t} \mathsf{mult} \mod N$, where $\mathsf{expn}\in\{0,1\}^{w_e}$ and $\mathsf{mult}\in \{0,1\}^{w_t}$. The cost of the unlookup is $L^{'}=\sqrt{(2^{w_t+w_e})}$ (See Figure~\ref{fig:unlookup-circuit}). The whole windowed lookup algorithm can be seen as a single nested loop, with the outer loop iterating over $w_e$ exponent qubits at a time, and the inner loop iterating over $w_t$ multiplication qubits at a time. Therefore, the total number of lookup additions to perform modular exponentiation is $\mathcal{O}\left(\frac{n^2}{w_t w_e}\right)$, with each of these lookup additions costing $\mathcal{O}\left(2^{w_e + w_t} + n\right)$ $\mathsf{Tof}$ gates, with the depth depending on the depth of the adder used (the $\mathcal{O}(n)$ comes from the linear cost of modular addition). Therefore, the final complexity of windowing-based modular exponentiation totals $\mathcal{O}\left(\frac{n^2}{w_t w_e}(2^{w_e + w_t} + n)\right)$ $\mathsf{Tof}$ gates. This complexity is minimized when $w_t=w_e = \frac{1}{2}\log n$, leading to an overall scaling of $\mathcal{O}\left(n^3/\log^2 n\right)$.  For this work, it is useful to look more carefully at the constant factors in the aforementioned complexity. Assuming a problem setting that involves the modular exponentiation of an $n$-bit modulus, with $n_e$ bits in the exponent, the $\mathsf{Tof}$ gate complexity of modular exponentiation is~\cite{gidney2021factor}:

$$
\#\mathsf{Tof} = \#\mathsf{LookupAdds}\times \mathsf{cost}(\mathsf{LookupAdd}).
$$
We know that the number of lookup additions can be calculated as the number of exponent windows times the number of multiplication windows. We also have a factor of 2 due to the fact that after a new exponent window is processed (using $\mathsf{LookupMul}$), the result of the previous exponent window must be reversibly uncomputed (using $\mathsf{LookupInvMul}$), thus giving us:
\begin{align*}
\#\mathsf{LookupAdds} = 2\frac{nn_e}{w_m w_e}.
\end{align*}
The cost of a single lookup-addition is made up of many components. First, we have $C_{\mathsf{Lookup}}$ the cost of lookuping a table of size ($2^{(w_e+w_m)}$), followed by the addition of this lookup value into a target register. For the addition, we will assume the target register is prepared in a coset state~\cite{gidney2019approximate}, hence allowing the use of Cuccaro's adder (instead of a circuit for modular addition) which results in a costs of $C_{\mathsf{ModAdd}}=2n+\mathcal{O}(\log n)$. The coset state register is padded by a number of qubits that scales logarithmically in the number total number of addition operations and desired fidelity~\cite{zalka_fast_1998,gidney2019approximate} (See Sec.~\ref{sec:preliminaries-background} for more details). In the context of windowed arithmetic, the number of qubits is $\log (\#\mathsf{LookupAdds}) \approx 2\log n + \log \epsilon$ where $\epsilon$ represents the desired error rate or fidelity relative to ideal additions. As a result, addition incurs an extra $\mathcal{O}(\log n)$ cost. Finally, we perform the unlookup (costing $C_{\mathsf{Unlookup}}$). Regardless of the initial set-up cost for the coset state, this approach becomes more convenient for this application, as circuits for modular addition --- despite our recent improvements~\cite{luongo2024measurement} --- have worse constant factors. One could potentially use the adder from\cite{gidney2018halving}, which uses $n$ $\mathsf{Tof}$ gates, but this construction requires $n$ extra ancilla qubits, hence we do not consider this possibility here.

The unlookup first involves a unary conversion on $w=\frac{1}{2}(w_e+w_m)$ bits, followed by a lookup over a table of $w$ address bits, ending finally with an uncomputation of the unary register. This gives us an overall cost of:
\begin{align*}
  \mathsf{cost}(\mathsf{LookupAdd})&=C_{\mathsf{Lookup}} + C_{\mathsf{ModAdd}} +C_{\mathsf{Unlookup}}\\
  &=(2^{w_e + w_m} + 2n + 3\sqrt{2^{w_e + w_m}}).  
\end{align*}
Combining the previous two expressions to compute the total $\mathsf{Tof}$ count gives us\footnote{As previously noted~\cite{gidney2019windowed}, if the unary conversion is performed using a temporary logical-$\mathsf{AND}$, we require no $\mathsf{Tof}$ gates for the unary uncomputation. Thus giving us 
    $\#\mathsf{Tof} = 2\frac{nn_e}{w_m w_e}(2^{w_e + w_m} + 2n + 2\sqrt{2^{w_e + w_m}})$
}
\begin{align*}
    \#\mathsf{Tof} = 2\frac{nn_e}{w_m w_e}(2^{w_e + w_m} + 2n + 3\sqrt{2^{w_e + w_m}})
\end{align*}
Similarly, we get a depth of
\begin{align*}
  \mathsf{depth}=2\frac{nn_e}{w_m w_e}(2^{w_e + w_m} + 2n + 3\sqrt{2^{w_e + w_m}}). 
\end{align*}
In summary, using the $\mathsf{QROM}$ lookup table construction of~\cite{babbush2018encoding}, the coset-state based adder of~\cite{zalka2006shor, gidney2019approximate}, and the unary-based uncomputation of~\cite{gidney2019windowed}, this results in the above mentioned $\mathsf{depth}$ and $\mathsf{Tof}$ counts for quantum modular exponentiation.

\section{Improvements}\label{sec:improvements}
In this section, we present several optimizations for windowed arithmetic circuits, particularly in scenarios involving multiple consecutive lookups. For cryptographically relevant ranges (i.e.\ when the number of exponent's bits are $1.5n$ and the modulo has $n$ bits) our combined improvements lead to a circuit size and depth reduction by $3\%$.  When the number of exponent qubits is $n$, the improvement goes up to $3.24\%$. 
We show how the cost of unlookups can be reduced by up to $66\%$ by deferring all unlookups to the end of each exponent window.
We also illustrate how certain addresses can be bypassed, reducing both circuit depth and the overall lookup cost. Furthermore, we demonstrate that by merging multiple lookup-addition operations into a single, larger lookup at the start of the modular exponentiation circuit, additional savings in both Toffoli gate count and circuit depth can be achieved. Finally, using existing techniques, we also show how the depth of unary conversion can be reduced.

\subsection{Deferred uncomputation}\label{subsec:imp1-deferred}

\begin{figure}[!ht]
    \centering
    \begin{subfigure}[b]{\textwidth} 
        \centering
        \includegraphics[width=\linewidth]{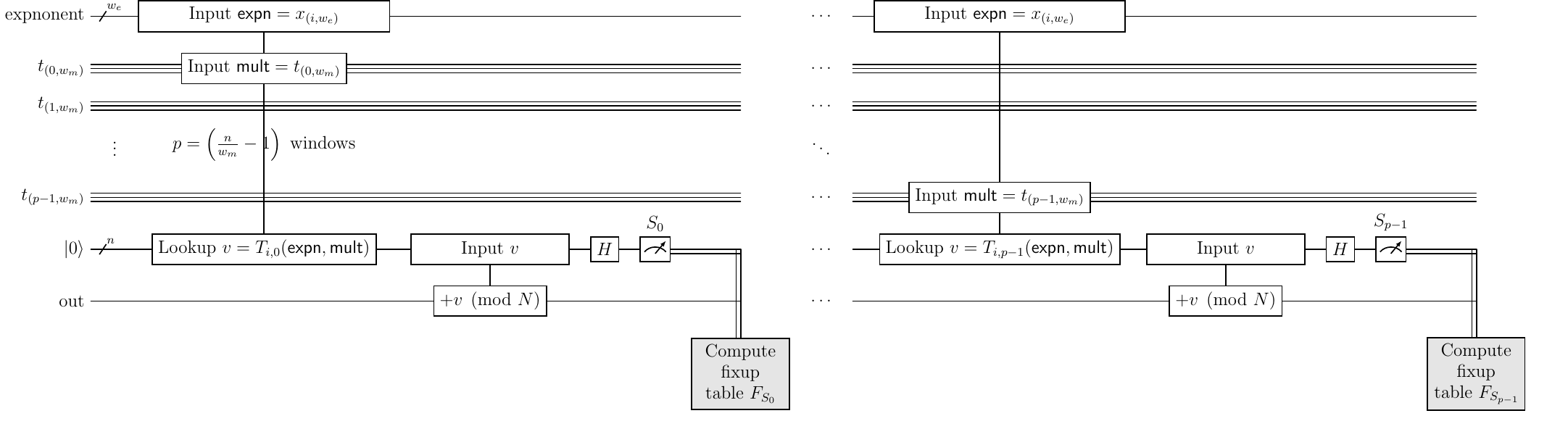}
    \end{subfigure}
    
    \vspace{10pt} 
    
    \begin{subfigure}[b]{\textwidth} 
        \centering
        \includegraphics[width=\linewidth]{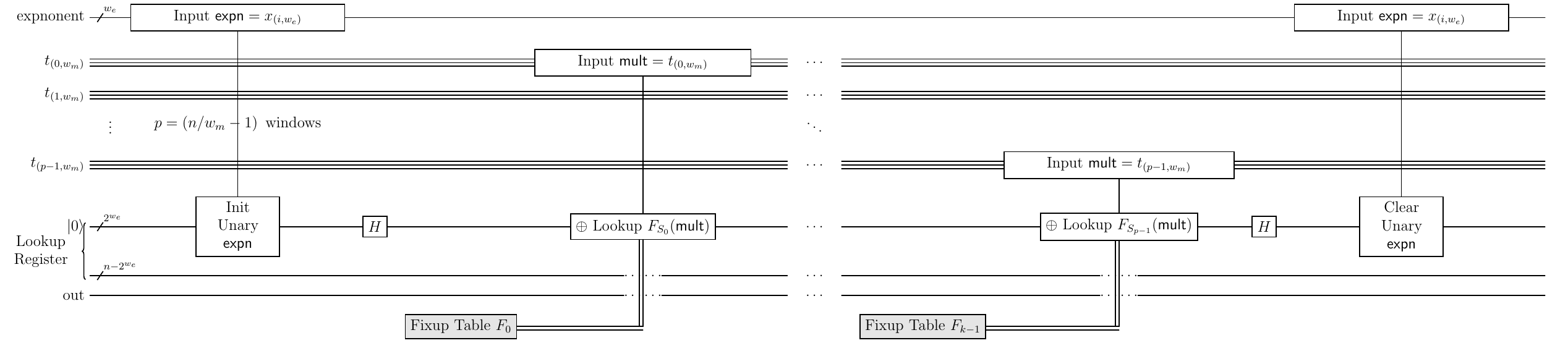}
    \end{subfigure}
    
    \caption{Our proposed circuit for deferred uncomputation. The lookup additions have 2 stages. In stage one, the windowed multiplication with lookup qubit reset and no phase correction (top) is performed, followed by the second stage, where the circuit for deferred phased correction (bottom) is executed. Notice that the unary conversion is performed only once, and on the exponent window as it is common for all lookups and can be reused. This is different from Figure~\ref{fig:windowing-original} where the unlookups or phase corrections are performed after every single lookup addition.}
    \label{fig:deferred-uncomputation-abstract}
\end{figure}
While unlookups are much cheaper than their corresponding lookups (as shown in~\cite{gidney2019windowed}), they can be made even cheaper in the multi-lookup setting, when part or the whole address register remains unchanged over many consecutive table lookups. This is the exact scenario in the case of the GE factoring algorithm, where for all the lookup-additions involved in multiplying $b^{x_{(i,w_e)}2^{iw_e}}$ (with index $i$, and window size $w_e$), the exponent window remains unchanged. For a fixed $i$, these lookups have address $a=x_{(i,w_e)}||t_{(j,w_m)}$ with $j\in\left\{0,1,\cdots,\frac{n}{w_m}-1\right\}$, where $t_{(j,w_m)}$ is a window over the multiplication qubits.

As mentioned before, the whole windowed lookup algorithm is a nested loop, with the outer loop iterating over $w_e$ exponent qubits at a time and an inner loop iterating over $w_t$ multiplication qubits at a time. Each unlookup is of the form:
{\small
\begin{align*}
    \left[\ket{x_{(i, w_e)}} \ket{t^{[i-1]}}_n \ket{\ell_{i,j}}_n\ket{r}_n \xmapsto{\mathsf{Unlookup}}\cdots \right]\equiv\left[
\begin{aligned}
    &\xmapsto{\mathsf{H},\,\mathsf{Measure}}(-1)^{S_j\cdot \ell_{i,j}}\ket{x_{(i, w_e)}} \ket{t^{[i-1]}}_n \ket{S_j}_n\ket{r}_n \\
    &\xmapsto{\mathsf{Reset}\text{ }S_j}(-1)^{S_j\cdot \ell_{i,j}}\ket{x_{(i, w_e)}} \ket{t^{[i-1]}}_n \ket{0}_n\ket{r}_n\\
    &\xmapsto{\mathsf{Unary}}(-1)^{S_j\cdot \ell_{i,j}}\ket{x_{(i, w_e)}} \ket{t^{[i-1]}}_n \ket{\mathsf{unary}(x_{(i, w_e)})}_{2^{w_e}}\ket{r}_n\ket{0}_{n-2^{w_e}}\\
    &\xmapsto{\mathsf{Lookup}\;F_{S_j}}\ket{x_{(i, w_e)}} \ket{t^{[i-1]}}_n \ket{\mathsf{unary}(x_{(i, w_e)})}_{2^{w_e}}\ket{r}_n\ket{0}_{n-2^{w_e}}\\
    &\xmapsto{\mathsf{Unary}^\dagger}\ket{x_{(i, w_e)}}\ket{t^{[i-1]}}_n \ket{0}_n\ket{r}_n\\
\end{aligned}
\right]
\end{align*}}%

Each lookup is addressed by $a=\mathsf{expn}||\mathsf{mult}$ that combines an (outer loop) exponentiation window index $\mathsf{expn}$ and an (inner loop) multiplication window index $\mathsf{mult}$. Gidney's \cite{gidney2019windowed} unary unlookup reduction is then applied to this construction.

 We improve this further, exploiting the nested loop structure. Notice that the outer index $\mathsf{expn}$ stays fixed while the circuit iterates over the inner index $\mathsf{mult}$. In our proposed circuit (Figure~\ref{fig:deferred-uncomputation-abstract}), we use $\mathsf{mult}$ as the address of the fixup table $F_{S_j}$ and the ``unarized'' $\mathsf{expn}$ as the target. The binary-to-unary initialization of $\mathsf{expn}$ needs to be done only once before entering the inner loop, and the erasure of the unary register only once right at the end of the inner loop\textemdash whereas, in the unmodified version, there is a reset and initialization after every iteration within the inner loop. Thus, the cost of our circuit is dominated by the phase lookups ($F_{S_j}$) within each iteration, with the one-time unary initialization and reset contributing negligibly in comparison. Overall, this leads to a halving of the Toffoli cost of unlookups in the original windowed modular exponentiation algorithm. After every lookup in the inner loop, let's instead perform the operation
\begin{align*}
    \left[\ket{x_{(i, w_e)}} \ket{t^{[i-1]}}_n \ket{\ell_{i,j}}_n\ket{r}_n \xmapsto{\mathsf{Def.~Unlookup}}\cdots \right]\equiv\left[
\begin{aligned}
    \xmapsto{\mathsf{H},~\mathsf{Measure}}&(-1)^{s_j\cdot \ell_{i,j}}\ket{x_{(i, w_e)}} \ket{t^{[i-1]}}_n \ket{S_j}_n\ket{r}_n \\
    \xmapsto{\mathsf{Reset~}S_j}&(-1)^{S_j\cdot \ell_{i,j}}\ket{x_{(i, w_e)}} \ket{t^{[i-1]}}_n \ket{0}_n\ket{r}_n\\
\end{aligned}
\right].
\end{align*}
The value $S_j$ is recorded classically, and a phase fixup table $F_{S_j}$ is constructed (indexed/addressed by $\mathsf{mult}$). At the end of the inner loop, we have $p=\left(\frac{n}{w_t}-1\right)$ classical bit strings $S_j$, and $p$ different phase fixup tables $F_{S_j}\;;\forall j\in \{0,1,\cdots,p-1\}$. The state at this stage is:
$$(-1)^{\sum_{k=0}^{p-1}S_k\cdot \ell_{i,j}}\ket{x_{(i, w_e)}} \ket{t^{[i-1]}}_n \ket{0}_n\ket{r}_n.$$
We can now construct a unary register with input $x_{(i,w_e)}$, and window over the register $t$ to perform our phase corrections with $F_{S_j}$, i.e.
\begin{align*}
(-1)^{\sum_{k=0}^{p-1}S_k\cdot \ell_{i,j}}\ket{x_{(i, w_e)}} &\ket{t^{[i-1]}}_n \ket{\mathsf{unary}(x_{(i,w_e)})}_{2^{w_e}}\ket{t^{[i]}}_n\ket{0}_{n-2^{w_e}}\\
&\xmapsto{\mathsf{Lookup~}F_{S_0}}(-1)^{\sum_{k=1}^{p-1}S_k\cdot \ell_{i,j}}\ket{x_{(i, w_e)}} \ket{t^{[i-1]}}_n \ket{\mathsf{unary}(x_{(i,w_e)})}_{2^{w_e}}\ket{t^{[i]}}_n\ket{0}_{n-2^{w_e}}\\
&\xmapsto{\mathsf{Lookup~}F_{S_1}}(-1)^{\sum_{k=2}^{p-1}S_k\cdot \ell_{i,j}}\ket{x_{(i, w_e)}} \ket{t^{[i-1]}}_n \ket{\mathsf{unary}(x_{(i,w_e)})}_{2^{w_e}}\ket{t^{[i]}}_n\ket{0}_{n-2^{w_e}}\\
&\;\;\vdots\\
&\xmapsto{\mathsf{Lookup~}F_{S_{p-2}}}(-1)^{S_{p-1}\cdot \ell_{i,{p-1}}}\ket{x_{(i, w_e)}} \ket{t^{[i-1]}}_n \ket{\mathsf{unary}(x_{(i,w_e)})}_{2^{w_e}}\ket{t^{[i]}}_n\ket{0}_{n-2^{w_e}}\\
&\xmapsto{\mathsf{Lookup~}F_{S_{p-1}}}\ket{x_{(i, w_e)}} \ket{t^{[i-1]}}_n \ket{\mathsf{unary}(x_{(i,w_e)})}_{2^{w_e}}\ket{t^{[i]}}_n\ket{0}_{n-2^{w_e}}\\
&\xmapsto{\mathsf{Unary^\dagger,~Swap}}\ket{x_{(i, w_e)}} \ket{t^{[i-1]}}_n \ket{0}_{n}\ket{t^{[i]}}_n.
\end{align*}
Here, $\mathsf{unary}^\dagger$ represents the uncomputation of the unary register. If we were to perform a more general complexity analysis on the impact of deferred uncomputation on windowed modular exponentiation, we see that
\begin{align*}
    \#\mathsf{Tof} & = 2\frac{nn_e}{w_m w_e}(2^{w_e + w_m} + 2n + 2\frac{w_m}{n}\cdot2^{w_e} + 2^{w_m}) \\
    \mathsf{depth} &=2\frac{nn_e}{w_m w_e}(2^{w_e + w_m} + 2n + 2\frac{w_m}{n}\cdot2^{w_e} + 2^{w_m}).
\end{align*}

\subsection{Selective lookups}\label{subsec:imp3-selective}
Although lookup tables differ for various moduli $N$ and exponent bases $b$, they share some structural similarities for specific addresses. During a lookup addition of the $i^{\text{th}}$ exponent window (of size $w_e$) and the $j^{\text{th}}$ multiplication window (of size $w_m$), we construct a classical table $T_{i,j}$ corresponding to addresses stored in the register 
$m_{(j,w_m)}||x_{(i,w_e)}$. For a classical bit string address $a=\mathsf{mult}||\mathsf{expn}$, with $\mathsf{mult} \in \{0,1,\cdots,2^{w_m}-1\}$ and $\mathsf{expn} \in \{0,1,\cdots,2^{w_e}-1\}$, the lookup table $T_{i,j}$ holds the following values:
\begin{align*}
T_{i,j}(\mathsf{mult},\mathsf{expn}):={ (b^{\mathsf{expn}2^{iw_e}} 2^{jw_m}\mathsf{mult}) \mod N}.
\end{align*}
$T_{i,j}$ contains $2^{w_e+w_m}$ distinct values (for all combinations of length $w_e$ and $w_m$ bit strings). 
Upon examining the values associated with addresses where $\mathsf{mult}=0$, we observe that $T_{i,j}(\mathsf{mult},*)=0$ regardless of $i$, $j$, and $\mathsf{expn}$. Consequently, we can initiate lookups directly from $\mathsf{mult}=1$ since there are no relevant values for $\mathsf{mult}=0$. This adjustment would save us $2^{w_m}$ lookups per query.

On the other hand, for all addresses with $\mathsf{expn} = 0$, we have $T_{i,j}(\mathsf{mult},\mathsf{expn}) = \mathsf{mult} 2^{jw_m}$. Thus, $\mathsf{mult}$ is directly copied to the target register. We propose to perform this copying at the start of every lookup. Note that the lookup table must be updated with new values: 
\begin{align*}
T'_{i,j}(\mathsf{mult},\mathsf{expn})=(b^{\mathsf{expn}2^{iw_e}} 2^{jw_m}\mathsf{mult} \mod N) \oplus 2^{j w_m}\mathsf{mult}.
\end{align*}
\begin{figure}[!ht]
    \centering
    \includegraphics[width=0.6\textwidth]{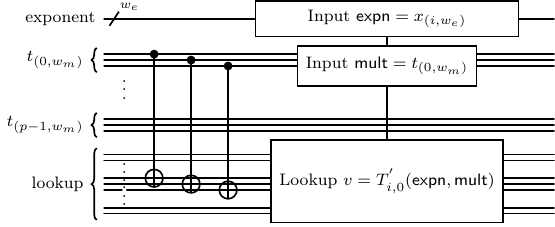}
    \caption{Proposed optimization for pruned table. The sequence of $\mathsf{CNOT}$s before the lookup operation, performs a copy operation into the lookup register. The lookup table is then modified to take into account the uncontrolled copy operation we initially performed. Here, $p$ represents the number of windows of size $w_m$ the register $t$ can be split into.}
    \label{fig:optimized_exp_zero_cnot_copy}
\end{figure}
This adjustment is necessary because the initial copying operation applies uniformly across all addresses. We need to correct the lookup values for cases where $\mathsf{expn} \neq 0$. The initial copying operation is equivalent to performing a lookup operation $v \leftarrow 2^{j w_m}\mathsf{mult}$. Subsequently, we perform a lookup of $T'_{i,j}(\mathsf{mult}, \mathsf{expn})$, which updates the lookup register as follows:
\begin{align*}
    v &\leftarrow 2^{jw_m}\mathsf{mult} \oplus T'_{i,j}(\mathsf{mult},\mathsf{expn}) \\
    &= 2^{jw_m}\mathsf{mult} \oplus (b^{\mathsf{expn}2^{iw_e}} 2^{jw_m}\mathsf{mult} \mod N) \oplus 2^{jw_m}\mathsf{mult} \\
    &= (b^{\mathsf{expn}2^{iw_e}} 2^{jw_m}\mathsf{mult}) \mod N \\
    &= T_{i,j}(\mathsf{mult},\mathsf{expn})
\end{align*}

This progression shows that the final value in the lookup register correctly corresponds to $T_{i,j}(\mathsf{mult},\mathsf{expn})$. This modification yields a saving of $2^{w_m}-1$ lookups (adjusting for the fact that otherwise, we would double-count the savings from the previous optimization for the case $\mathsf{mult}=0$). Refer to Fig~\ref{fig:optimized_exp_zero_cnot_copy} for a detailed circuit diagram of this copy-and-update-table lookup operation.

Analytically, we see that the size and depth complexity of the circuit becomes
\begin{align*}
    \#\mathsf{Tof} & = 2\frac{nn_e}{w_m w_e}(2^{w_e + w_m} + 2n + 3\sqrt{2^{w_e + w_m}} - 2^{w_e}) \\
    \mathsf{depth} &=2\frac{nn_e}{w_m w_e}(2^{w_e + w_m} + 2n + 3\sqrt{2^{w_e + w_m}} - 2^{w_e}).
\end{align*}

\subsection{Larger initial lookup}\label{subsec:imp4-larger}
While windowed modular exponentiation typically processes the exponent in small windows of size $\frac{1}{2}\log n$ to balance precomputation and the number of iterations of the lookup-additions, we investigate an optimization for the algorithm's beginning phase. By combining multiple initial exponent windows into a single, larger lookup operation, we show below how to slightly reduce the number of modular multiplications and thus decrease the overall Toffoli gate count. Recall that with \( x_{(i, w_e)} \) we  represent the \( i \)-th window of size \( w_e \) in the exponent and with  \( t_{(j, w_t)} \)  we represent the \( j \)-th window of size \( w_t \) in the multiplication register.
We also recall from Section~\ref{subsec:windowing} how modular exponentiation is performed using windowing:

\begin{align*}
\ket{x_{(0,w_e)}}_{w_e}\ket{1}_{n}\ket{0}_{2n}\xmapsto{\times b^{x_{(0, w_e)}} \mod N}&\ket{x_{(0,w_e)}}_{w_e}\ket{t^{[0]}}_n\ket{0}_{2n} \nonumber \\
\ket{x_{(1,w_e)}}_{w_e}\ket{t^{[0]}}_n\ket{0}_{2n}\xmapsto{\times b^{2^{w_e} x_{(1, w_e)}} \mod N}&\ket{x_{(1,w_e)}}_{w_e}\ket{t^{[1]}}_n\ket{0}_{2n}\nonumber  \\
&\vdots \text{ Iterate over $i$ exponent windows} \nonumber \\
\ket{x_{(i,w_e)}}_{w_e}\ket{t^{[i-1]}}_n\ket{0}_{2n}\xmapsto{\times b^{\left(2^{i \cdot w_e}\right) x_{(i, w_e)}} \mod N}&\ket{x_{(i,w_e)}}_{w_e}\ket{t^{[i]}}_n\ket{0}_{2n}\\
&\vdots \text{ Iterate over the rest of the exponent windows} \nonumber \\
\ket{x_{(k-1,w_e)}}_{w_e}\ket{t^{[k-2]}}_n\ket{0}_{2n}\xmapsto{\times b^{\left(2^{(k-1) \cdot w_e}\right) x_{(k-1, w_e)}} \mod N}&\ket{x_{(k-1,w_e)}}_{w_e}\ket{t^{[k-1]}}_n\ket{0}_{2n}
\end{align*}

\begin{figure}[!ht]
    \centering
    \includegraphics[width=\textwidth]{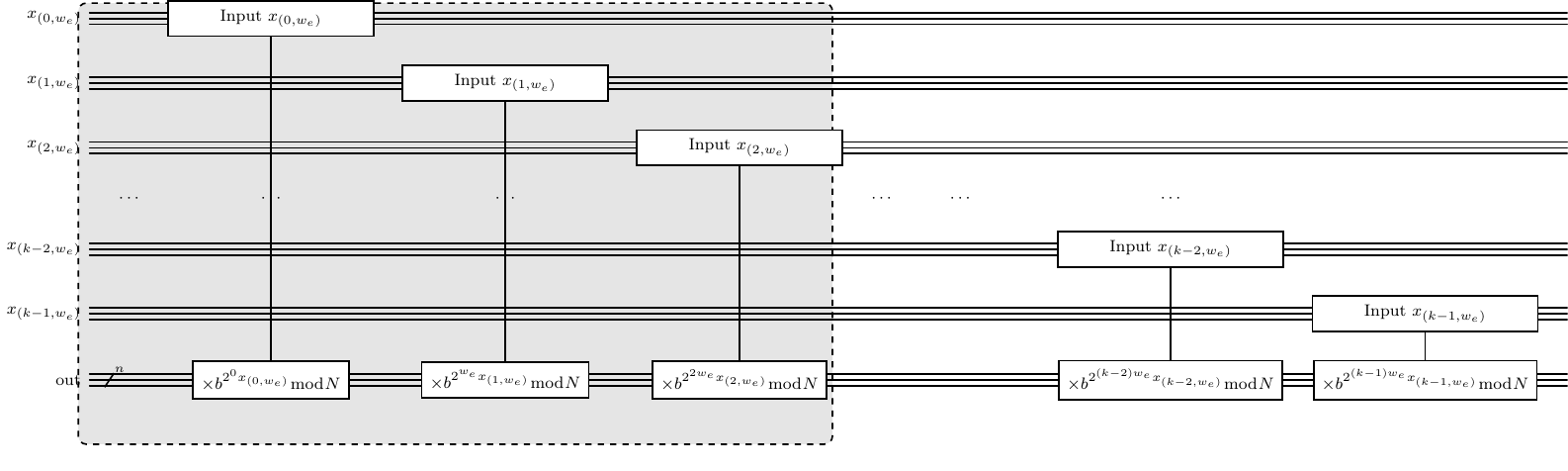}
    \caption{Standard windowed modular exponentiation. The first three exponent windows $(x_{(0,w_e)}, x_{(1,w_e)}, \text{and } x_{(2,w_e)})$ are highlighted, demonstrating the typical approach of processing the exponent in small windows. Each window leads to a separate modular multiplication operation}
    \label{fig:larger-initial-lookup-part1}
\end{figure}
\begin{figure}[!ht]
    \centering
    \includegraphics[width=0.75\textwidth]{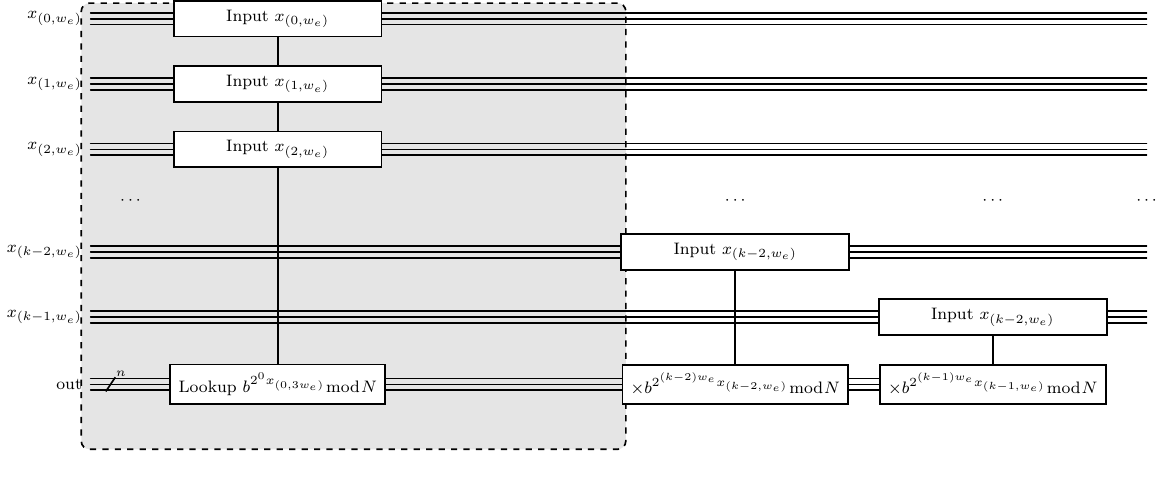}
    \caption{Optimized initial stage of quantum modular exponentiation. In contrast to the standard approach shown in Figure~\ref{fig:larger-initial-lookup-part1}, the first three exponent windows have been combined into a single, larger lookup operation. This optimization reduces the number of modular multiplications required in the early stages of the algorithm, leading to savings in $\Tof$ count and depth.}
    \label{fig:larger-initial-lookup-part2}
\end{figure} 
The initial sequence of modular multiplications in the modular exponentiation algorithm can be reduced to a single lookup operation over the first $(i+1)w_e$ bits of the exponent\footnote{Note, in these set of equations, we only show the current exponent window of qubits that are involved in the windowing operation. Based on the implementation of modular exponentiation, the previous exponent windows either remain in memory until all windows are exponentiated, or are repeatedly measured and reused using a Semiclassical QFT~\cite{griffiths1996semiclassical}}. However, as $i$ increases, this approach becomes more expensive than regular windowed modular multiplication, potentially negating the benefits of windowed arithmetic due to exponential growth in lookup table size. We explore the tradeoff between a exponentiation based on direct lookups (for the first $e$ bits of the exponent) versus the cost of exponentiation based on lookup-additions.
To determine the optimal size of this initial ``large'' lookup, we compare its cost to that of performing $2i$ separate windowed multiplications (one for the out-of-place multiplication register and one for uncomputation of the previous multiplicand; see Section~\ref{subsec:windowing}). From Figure~\ref{fig:direct-exp-vs-lookup-add-exp}, for the problem of $2048$-bit modular exponentiation, we see that it only costs about $1$ million $\Tof$ gates, versus $10$ million $\Tof$ gates for the same operation to be performed with a window size of $5$ (i.e.\ using $5$ bit exponent windows). More specifics on these costs calculations can be seen in Appendix.~\ref{apx:larger-initial-lookup}

If we were to perform a more general complexity analysis (e.g.\ looking up $n'_e$ exponent bits directly) on the impact of a larger initial lookup on windowed modular exponentiation, we see that
\begin{align*}
    \#\mathsf{Tof} & =2^{n'_e} + 2\frac{n(n_e-n'_e)}{w_m w_e}(2^{w_e + w_m} + 2n + 3\sqrt{2^{w_e + w_m}}), \\
    \mathsf{depth} &=2^{n'_e} + 2\frac{n(n_e-n'_e)}{w_m w_e}(2^{w_e + w_m} + 2n + 3\sqrt{2^{w_e + w_m}}).
\end{align*}

\subsection{Lower-depth unary conversion}\label{subsec:imp5-lower-depth-unary}

\begin{figure}[!ht]
    \centering
    \includegraphics[width=0.5\textwidth]{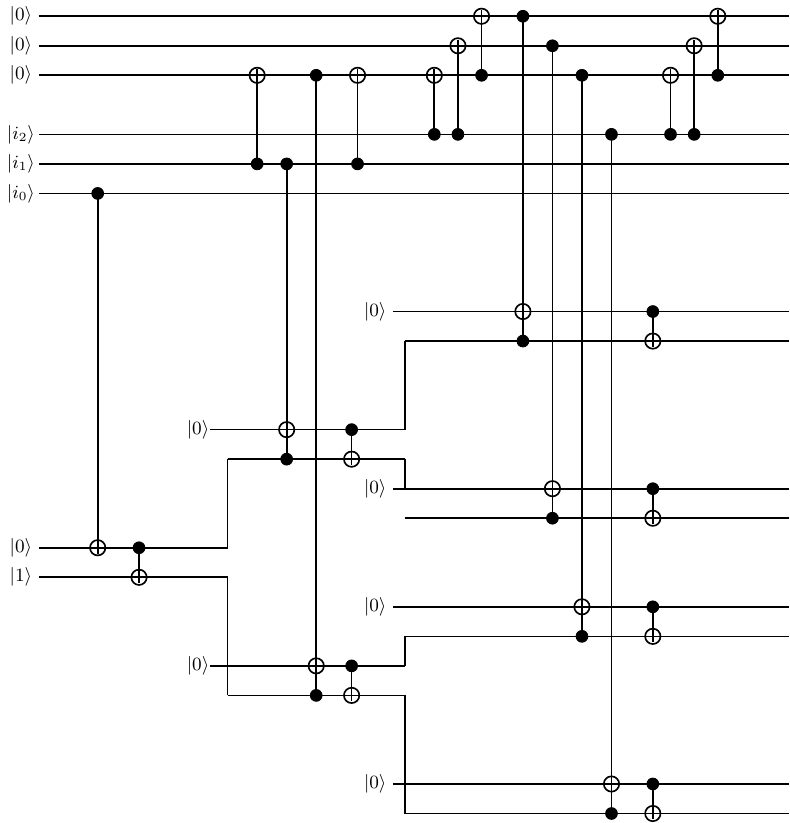}
    \vspace{0.45cm}
    \caption{Illustration of the address routing stage of a bucket brigade $\mathsf{QRAM}$ with $n=3$ address qubits. This figure here is a part of the architecture proposed by ~\cite{arunachalam2015robustness} and edited in \cite{doriguello2024practicality}}
    \label{fig:unar-conversion-address-routing-bb}
\end{figure}

In the context of uncomputing quantum table lookups, the unary conversion step plays a pivotal role. Typically, this involves taking half of the address qubits and applying a unary conversion circuit, as illustrated in Figure~\ref{fig:unary_conversion}. However, the depth of this standard unary conversion circuit grows exponentially with the input size. Specifically, for an address register of size \( k \) qubits, using half of these qubits for unary conversion results in a $\mathsf{Tof}$ depth of \( 2^{k/2} - 1 \). To achieve lower depth in unary conversion, we draw on concepts from quantum random access memory ($\mathsf{QRAM}$) architectures. $\mathsf{QRAM}$ provides a means to reduce the depth of lookup operations, albeit at the cost of an exponential increase in qubit usage relative to the address size. A $\mathsf{QRAM}$ lookup can be decomposed into two main stages: the address routing stage and the lookup stage. Our focus is on the address routing stage (as illustrated in Figure~\ref{fig:unar-conversion-address-routing-bb}), which involves a combination of address register fanout and unary encoding. This stage typically requires approximately $2^w$ qubits, where $w$ is the size of the address register. The circuit can be seen as traversing down a binary tree, and setting the leaf at index $i$ to $1$, if and only if $a=i$. All the other qubits are set to $0$. This is exactly what a $\mathsf{unary}$ encoding does. This method of unary encoding has a $\mathsf{Tof}$ depth of $w$ for a $w$ bit address, with a space cost coming from the largest address fanout and space required to store the unary register. This gives us a total space cost of $2^{w} + 2^{w-1} - 1$. Finally, the $\mathsf{Tof}$ count is $2^w-1$. We have an improvement in the depth compared to the original unary circuit, ours being $w$ vs.\ the original $2^w-1$. The price we pay is the space complexity, ours being $2^{w}+2^{w-1}-1$ while the original is $2^w$.

\FloatBarrier

In the specific context of factoring problems or modular exponentiation, the number of qubits available for lookup operations is typically exponential in the number of address qubits. The size and depth complexities of the circuit become
\begin{align*}
    \#\mathsf{Tof} & = 2\frac{nn_e}{w_m w_e}(2^{w_e + w_m} + 2n + 3\sqrt{2^{w_e + w_m}}), \\
    \mathsf{depth} &=2\frac{nn_e}{w_m w_e}(2^{w_e + w_m} + 2n + 2\sqrt{2^{w_e + w_m}} + w_e-1).
\end{align*}

\section{Measuring the impact on attacks against \texorpdfstring{$\mathsf{RSA}$}{RSA}}\label{sec:application-to-break-rsa} 
Many quantum algorithms have been proposed that offer an asymptotic speedup over classical counterparts. However, practical implementation requires careful consideration of constant factors, which includes the challenge of estimating the non-asymptotic cost in terms of time and space. A significant overhead in running these algorithms is the number of physical qubits needed to protect logical information from noise during computation. 

With a clearer understanding of the parameters for large-scale quantum computers—such as qubit modality, expected gate error rates, gate speeds, and potential quantum error-correcting codes (QECC)—estimating the fault-tolerant costs associated with specific quantum algorithms has become more feasible. A given algorithm does not have a unique compilation; therefore, optimizing these compilations for resource efficiency is among the most critical and complex problems in fault-tolerant quantum computing. Resource estimation plays a vital role in addressing this challenge.

In this section, we build upon the resource estimation framework introduced by Gidney and Eker\aa~\cite{gidney2021factor} by incorporating our proposed enhancements to windowed arithmetic. Our analysis targets a superconducting qubit architecture, with computations encoded in rotated surface codes and logical operations performed using lattice surgery~\cite{horsman2012surface, fowler2018low}. We assume a physical error rate of \(10^{-3}\) (and also test with a more optimistic \(10^{-4}\) error rate), reflecting the fidelity of operations such as single- and two-qubit gates, state initialization, and measurements. Additionally, we consider a cycle time of 1$\mu$s, representing the duration required to measure all stabilizers of a surface code patch. These parameters are chosen to align with the anticipated capabilities of future fault-tolerant quantum hardware and provide a consistent baseline for comparison. For a distance-\(d\) rotated surface code, we require \(2(d+1)^2\) physical qubits to encode a single logical qubit of information. To compare with~\cite{gidney2021factor}, our goal is to minimize the \emph{skewed volume}, a metric that balances the tradeoffs between qubit count and runtime, defined as:
\[
\text{skewed\_volume} = (\text{Mqb})^{q} \times \mathbb{E}[\text{hrs}]
\]
where Mqb is the total number of physical qubits in megaqubits (millions of qubits), and \(\mathbb{E}[\text{hrs}]\) is the expected runtime in hours, accounting for the likelihood of retries due to logical errors. The exponent \( q \) reflects the tradeoff between qubit resources and runtime, and in~\cite{gidney2021factor}, \( q \) is taken to be $1.2$, indicating a preference for reducing the physical qubit count (Mqb) over the expected runtime (\(\mathbb{E}[\text{hrs}]\)). The value of \( q \) can be adjusted depending on the use case.

\begin{table}[htbp]
\centering
\begin{adjustbox}{width=\textwidth}
\begin{tabular}{c|c|c|c|c|c|c|c|c||c|c|c|c|c|c|c}
$\bm{n}$ & $\bm{n_e}$ & \textbf{gate err} & $L_1$ & $L_2$ & $d_{\mathrm{off}}$ & $g_{\mathrm{mul}}$ & $g_{\mathrm{exp}}$ & $g_{\mathrm{sep}}$ & \% & v.p.r & $\mathbb{E}[\text{vol}]$ & Mqb & hrs & $\mathbb{E}[\text{hrs}]$ & B Tofs \\
\hline
$1024$ & $1493$ & $10^{-3}$ & $15$ & $27$ & $5$ & $5$ & $5$ & $1024$ & $6\%$ & $0.507$ & $0.539$ & $9.624$ & $1.264$ & $1.344$ & $0.407$ \\
$2048$ & $3029$ & $10^{-3}$ & $15$ & $27$ & $4$ & $5$ & $5$ & $1024$ & $31\%$ & $4.047$ & $5.865$ & $19.249$ & $5.046$ & $7.313$ & $2.698$ \\
$3072$ & $4565$ & $10^{-3}$ & $17$ & $29$ & $6$ & $4$ & $5$ & $1024$ & $9\%$ & $18.328$ & $20.141$ & $37.897$ & $11.607$ & $12.755$ & $9.885$ \\
$4096$ & $6101$ & $10^{-3}$ & $17$ & $31$ & $9$ & $4$ & $5$ & $1024$ & $5\%$ & $47.963$ & $50.488$ & $54.616$ & $21.077$ & $22.186$ & $23.038$ \\
$1024$ & $1493$ & $10^{-4}$ & $7$ & $13$ & $4$ & $5$ & $5$ & $512$ & $5\%$ & $0.07$ & $0.073$ & $2.637$ & $0.633$ & $0.667$ & $0.415$ \\
$2048$ & $3029$ & $10^{-4}$ & $7$ & $13$ & $4$ & $5$ & $5$ & $512$ & $21\%$ & $0.558$ & $0.706$ & $5.273$ & $2.538$ & $3.212$ & $2.774$ \\
$3072$ & $4565$ & $10^{-4}$ & $9$ & $15$ & $5$ & $5$ & $5$ & $768$ & $2\%$ & $2.92$ & $2.98$ & $9.12$ & $7.686$ & $7.843$ & $8.595$ \\
$4096$ & $6101$ & $10^{-4}$ & $9$ & $15$ & $5$ & $4$ & $5$ & $512$ & $3\%$ & $6.791$ & $7.001$ & $15.241$ & $10.693$ & $11.024$ & $23.773$ \\
\end{tabular}
\end{adjustbox}
\begin{adjustbox}{width=\textwidth}
\begin{tabular}{c|c|c|c|c|c|c|c|c||c|c|c|c|c|c|c}
$\bm{n}$ & $\bm{n_e}$ & \textbf{gate err} & $L_1$ & $L_2$ & $d_{\mathrm{off}}$ & $g_{\mathrm{mul}}$ & $g_{\mathrm{exp}}$ & $g_{\mathrm{sep}}$ & \% & v.p.r & $\mathbb{E}[\text{vol}]$ & Mqb & hrs & $\mathbb{E}[\text{hrs}]$ & B Tofs\\
\hline
$1024$ & $1493$ & $10^{-3}$ & $15$ & $27$ & $4$ & $5$ & $5$ & $1024$ & $6\%$ & $0.488$ & $0.519$ & $9.624$ & $1.217$ & \textcolor{blue}{1.295} & \textcolor{blue}{$0.393$} \\
$2048$ & $3029$ & $10^{-3}$ & $17$ & $27$ & $6$ & $5$ & $5$ & $1024$ & $20\%$ & $4.419$ & $5.524$ & \textcolor{red}{21.616} & $4.906$ & \textcolor{blue}{6.133} & \textcolor{blue}{$2.656$} \\
$3072$ & $4565$ & $10^{-3}$ & $17$ & $29$ & $4$ & $4$ & $5$ & $1024$ & $9\%$ & $17.727$ & $19.48$ & $37.897$ & $11.226$ & \textcolor{blue}{12.336} & \textcolor{blue}{$9.742$} \\
$4096$ & $6101$ & $10^{-3}$ & $17$ & $31$ & $8$ & $4$ & $5$ & $1024$ & $5\%$ & $46.424$ & $48.867$ & $54.616$ & $20.4$ & \textcolor{blue}{21.474} & \textcolor{blue}{$22.8$} \\
$1024$ & $1493$ & $10^{-4}$ & $7$ & $13$ & $4$ & $5$ & $5$ & $512$ & $5\%$ & $0.067$ & $0.071$ & $2.637$ & $0.612$ & \textcolor{blue}{0.644} & \textcolor{blue}{$0.402$} \\
$2048$ & $3029$ & $10^{-4}$ & $7$ & $13$ & $3$ & $5$ & $5$ & $512$ & $21\%$ & $0.541$ & $0.684$ & $5.273$ & $2.461$ & \textcolor{blue}{3.115} & \textcolor{blue}{$2.72$} \\
$3072$ & $4565$ & $10^{-4}$ & $9$ & $15$ & $5$ & $5$ & $5$ & $768$ & $2\%$ & $2.851$ & $2.909$ & $9.12$ & $7.503$ & \textcolor{blue}{7.656} & \textcolor{blue}{$8.486$} \\
$4096$ & $6101$ & $10^{-4}$ & $9$ & $15$ & $5$ & $4$ & $5$ & $512$ & $3\%$ & $6.588$ & $6.792$ & $15.241$ & $10.374$ & \textcolor{blue}{$10.695$} &\textcolor{blue}{ $23.559$} \\
\end{tabular}
\end{adjustbox}
\caption{The columns represent: the number of bits of the numbers to factor ($n$), the bits of the exponentiation register ($n_e$), the gate error (\textbf{gate err}), the level 1 and level 2 distances for the CCZ factories ($L_1$ and $L_2$), the deviation of the padding ($\text{d}_{\mathrm{off}}$) related to the error introduced by coset representation, the window size for the multiplication register of windowed arithmetic ($g_{\mathrm{mul}}$), the window size for the exponent register of windowed arithmetic ($g_{\mathrm{exp}}$), the size of the adder pieces for the oblivious carry runways ($g_{\mathrm{sep}}$), the probability of having an error in the computation (retry risk - \%), volume of the computation for a single run expressed in megaqubit-days (v.p.r.), expected volume of the computation taking into account the retry risk ($\mathbb{E}[\text{volume}]$), number of megaqubits (Mqb, $10^6$), runtime in hours for a single run (hours), expected runtime taking into account the error ($\mathbb{E}[\text{hrs}]$), and number of $\Tof$ gates in billions (B Tofs). Values in \textcolor{red}{red} indicate degradation (increases) and values in \textcolor{blue}{blue} indicate improvements (decreases) compared to the corresponding values in the first table.}\label{tab:impact_qecc_superconducting}\end{table}

\begin{table}[ht]
\centering
\begin{adjustbox}{width=0.85\textwidth}
\begin{tabular}{c|c|c|c|c|c|c|c|c|c}
\hline
$\bm{n}$ & $\bm{n}_e$ & \textbf{gate err} & Mqb & \multicolumn{2}{c|}{$\mathbb{E}[\text{hrs}]$} & \% improvement & \multicolumn{2}{c|}{B Tofs} & \% improvement \\
\cline{5-6} \cline{8-9}
& & & & GE & Ours & $\mathbb{E}[\text{hrs}]$ & GE & Ours & in B Tofs \\
\hline
2048 & 3029 & $10^{-3}$& 14.747 & 12.361 & 11.912 & 3.63\% & 2.656 & 2.607 & 1.85\% \\
2048 & 3029 & $10^{-3}$& 15.592 & 10.584 & 10.332 & 2.39\% & 2.665 & 2.609 & 2.11\% \\
2048 & 3029 & $10^{-3}$& 17.492 & 9.834 & 9.548 & 2.91\% & 2.656 & 2.621 & 1.31\% \\
2048 & 3029 & $10^{-3}$& 18.513 & 8.428 & 8.242 & 2.22\% & 2.665 & 2.616 & 1.85\% \\
2048 & 3029 & $10^{-3}$& 19.249 & 7.313 & 7.08 & 3.17\% & 2.698 & 2.643 & 2.06\% \\
2048 & 3029 & $10^{-3}$& 21.616 & 6.388 & 6.133 & 4.01\% & 2.698 & 2.656 & 1.57\% \\
2048 & 3029 & $10^{-3}$& 24.001 & 6.334 & 6.084 & 3.98\% & 2.695 & 2.656 & 1.45\% \\
2048 & 3029 & $10^{-3}$& 25.265 & 5.367 & 5.192 & 3.25\% & 3.041 & 2.988 & 1.76\% \\
2048 & 3029 & $10^{-3}$& 27.308 & 5.364 & 5.181 & 3.39\% & 3.045 & 2.992 & 1.75\% \\
2048 & 3029 & $10^{-3}$& 29.184 & 4.772 & 4.497 & 5.76\% & 3.125 & 3.079 & 1.48\% \\
2048 & 3029 & $10^{-3}$& 32.602 & 4.012 & 3.79 & 5.55\% & 3.132 & 3.085 & 1.51\% \\
2048 & 3029 & $10^{-3}$& 40.075 & 3.478 & 3.329 & 4.29\% & 3.718 & 3.085 & 17.02\% \\
\hline
\end{tabular}
\end{adjustbox}
\caption{A summary of improvements from Table~\ref{tab:impact_qecc_superconducting_increasing_qubit_count} (reported in the Appendix). In Table~\ref{tab:impact_qecc_superconducting_increasing_qubit_count} we study the lowest expected runtimes of the two algorithms for attacking $\mathsf{RSA}$-$2048$ keys with increasing qubit counts for a given number of physical qubits with (\textbf{gate err} $10^{-3}$). Every row in this table compares the improvements for the expected runtime and $\Tof$ count (in billions)  for a fixed number of physical qubits (Mqb).}
\label{tab:comparing_runtime_and_btofs_improvements}
\end{table}

\begin{figure}[H]
    \centering
    \includegraphics[width=0.9\textwidth]{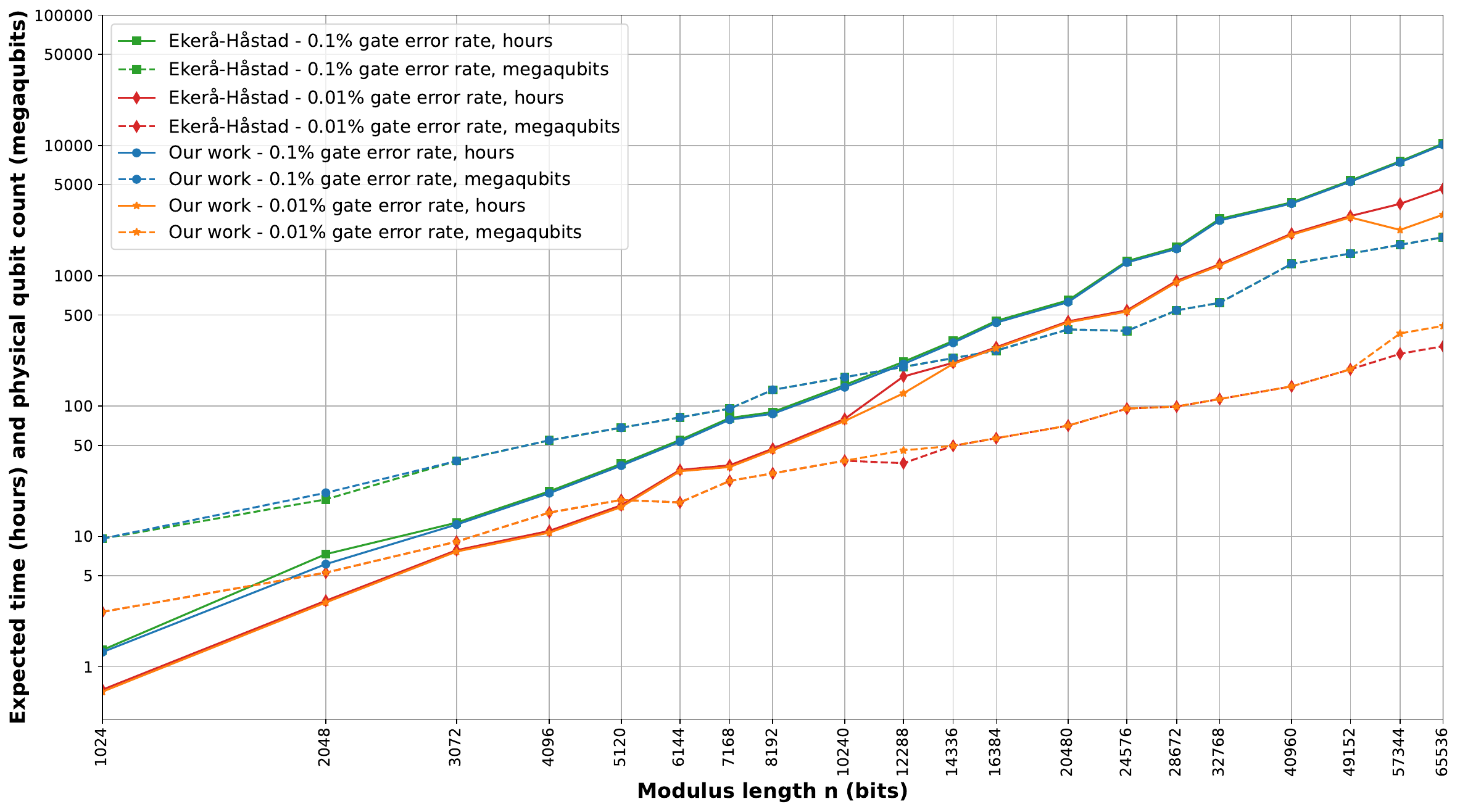}
     \caption{Scaling comparison between the windowing \cite{gidney2021factor} and optimized windowing (this work) of space and time costs with $10^{-3}$ and $10^{-4}$ gate error rates for large $\mathsf{RSA}$ key sizes.}    \label{fig:simpleimprovements-comparison-all-key-sizes}
\end{figure}

The improvements\footnote{To generate the graphs and the table, we used~\url{https://github.com/Inveriant/TAMARIND} which is based on the code of~\cite{gidney2021factor}} described in Sections~\ref{subsec:imp1-deferred}, \ref{subsec:imp3-selective}, \ref{subsec:imp4-larger} and \ref{subsec:imp5-lower-depth-unary} focus on reducing the $\Tof$ count and computational volume for modular arithmetic operations. When combined, these optimizations yield reductions in the $\Tof$ count and depth for cryptographically relevant attacks on $\mathsf{RSA}$ factoring by $1.5\%$ to $3.4\%$, as summarized in Table~\ref{tab:impact_qecc_superconducting}. More specifically, in the second line of the bottom part of the Table, we see nearly a $16\%$ reduction in the expected runtime of the algorithm for factoring $\mathsf{RSA}$-$2048$ bit integers at the cost of a $12\%$ increase in the physical qubit count (owing to the reduced retry risk of the algorithm) when comparing the lowest skewed volume estimates of GE and ours. However, this increase in qubit count does not paint the full picture because the computed estimates are an artifact of the metric we are trying to optimize (the skewed volume in our case). The GE algorithm can also achieve lower expected runtimes at the cost of more physical qubits. For a more fair comparison, in Table~\ref{tab:comparing_runtime_and_btofs_improvements} we explore the lowest achievable runtimes of both algorithms for a given number of physical qubits. We notice a $2$--$6\%$ reduction in expected runtime when incorporating our proposed improvements.  More tradeoffs are explored in Appendix~\ref{apx:mqb_tradeoffs_ge_v_ours}. In Figure~\ref{fig:rsa-256-bit} and Figure~\ref{fig:simpleimprovements-comparison-all-key-sizes} we provide a broader comparison across various $\mathsf{RSA}$ key sizes.

\begin{figure}[H]
\centering
    \includegraphics[width=0.9\textwidth]{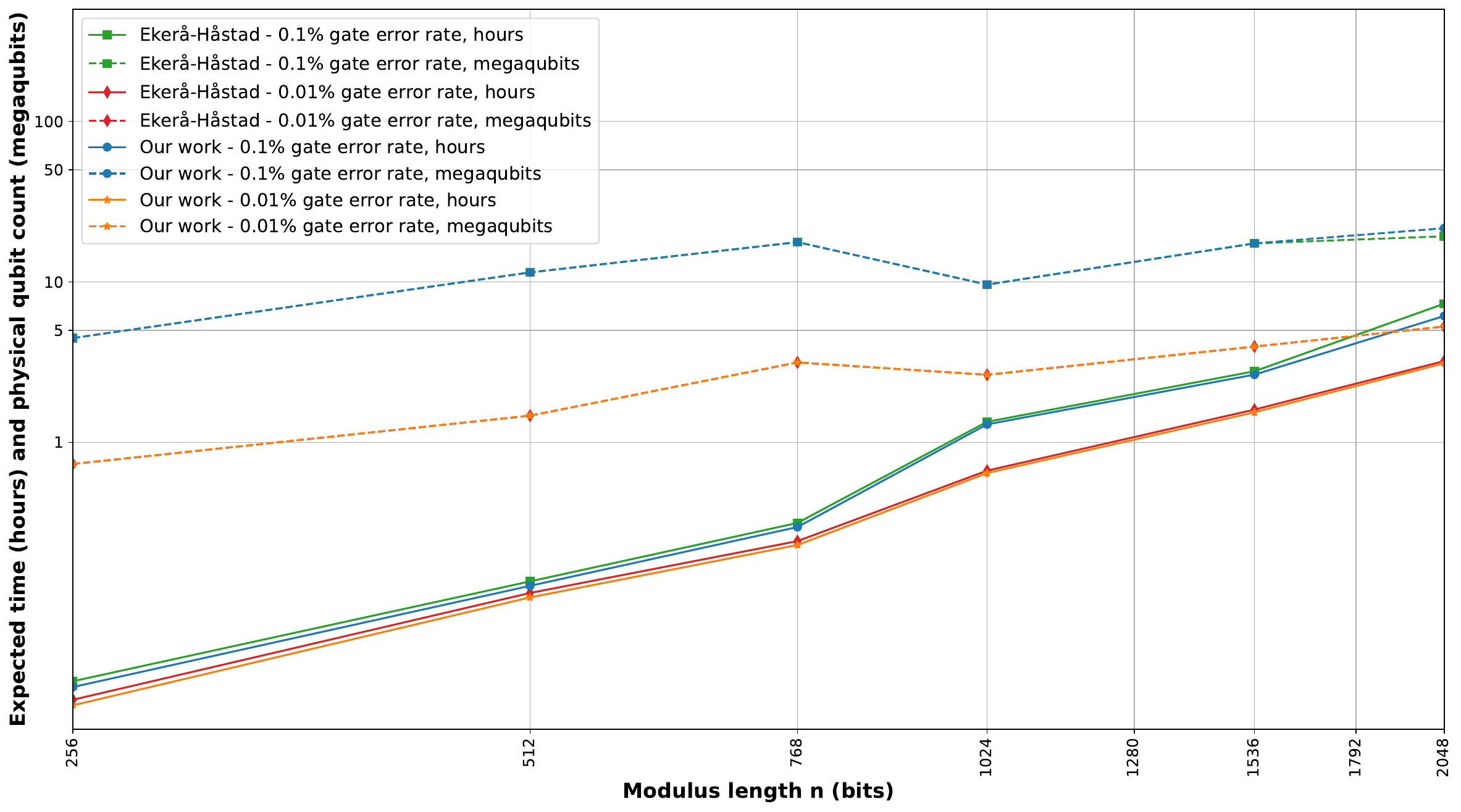}
    \caption{Scaling comparison between the windowing \cite{gidney2021factor} and optimized windowing (this work) of space and time costs with $10^{-3}$ and $10^{-4}$ gate error rates for small $\mathsf{RSA}$ key sizes.}
    \label{fig:rsa-256-bit}
\end{figure}

\section{Discussion and conclusions}\label{sec:discussion}
In this paper, we examine Gidney's windowed arithmetic construction \cite{gidney2019windowed} for efficient modular multiplication and explore how it can be further optimized to reduce resource estimates for fault-tolerant implementations. Through the improvements described in this work, the quantum factoring circuit using windowed arithmetic can be made (slightly) cheaper on the algorithmic level. In Section~\ref{sec:improvements}, we present four algorithmic optimizations to the windowed arithmetic method, focusing on reducing the $\Tof$ cost and depth of lookups and unlookups, as well as minimizing the number of required lookups in the modular exponentiation algorithm used by Gidney. Finally, in Section~\ref{sec:application-to-break-rsa} we test these proposed improvements within Gidney--Eker\aa's resource estimation framework \cite{gidney2021factor} and demonstrate a reduction in the overall computational volume for factoring $\mathsf{RSA}$-$2048$ integers.

\paragraph{Recent improvements on factoring.}
While more recent factoring algorithms~\cite{regev2023efficient,chevignard2024reducing} (which we discussed in Section~\ref{sec:variants-of-quantum-factoring-alg}) present exciting theoretical advancements, they are still in their nascent stages, with considerable optimization required for practical execution. Some more recent results have made improvements to Regev's factoring algorithm~\cite{ragavan2024regev, ragavan2023optimizing, ekeraa2024extending}. To our knowledge, the best implementation of Regev's algorithm~\cite{ragavan2024regev} uses only $10.4n$ qubits --- vs.\ $3n$ qubits for the GE algorithm --- and translates to a number of physical resources higher than the ones required by GE's algorithm. The work of~\cite{chevignard2024reducing} --- under some heuristic assumptions --- requires only $\frac{n}{2} + o(1)$ qubits (specifically $1730$ for $\mathsf{RSA}$-$2048$), but requires $6.9\times 10^{10}$ $\mathsf{Tof}$ gates, which is nearly $25$ times the number of $\mathsf{Tof}$ gates required by GE's algorithm (or our improvements). Furthermore, they require an average of $40$ repetitions to succeed. We anticipate that new algorithmic subroutines will emerge in the near future, further reducing the physical resource requirements. In contrast,  windowing arithmetic offers a more practical avenue for immediate improvements. Therefore, in this paper, we focus on optimizing the windowed arithmetic circuits in the GE algorithm and quantifying the impact of these improvements on the costs of breaking $\mathsf{RSA}$ and similar cryptographic protocols using quantum computers.

\paragraph{Further techniques for reducing the depth.}
Several quantum lookup table architectures exist, including parallel methods that trade size for reduced depth. These architectures have been explored in various works~\cite{giovannetti2008architectures,jaques2023qram,allcock2023constant,low2024trading,yuan2023optimal,zhu2024unified}. The bucket brigade QRAM is notable for achieving favorable query fidelities but requires a large qubit count that scales with both the number of memory elements and the word length ($n$) of each element~\cite{hann2021practicality}  For memory lookups and addition, using such architectures could reduce the depth of a $\mathsf{LookupAdd}$ operation from $(2^{w_e + w_m} + 2n + 3\sqrt{2^{w_e + w_m}} )$ to $\mathcal{O}(\frac{n}{k}(w'_e + w'_m))$, with a space overhead of $\mathcal{O}(n + k2^{w'_e + w'_m})$ for $1\leq k \leq n$. However, as mentioned previously, this comes at a cost: when $k=1$, the QRAM is sequential and depth becomes linear in the word length (which for us is $n$), losing the tradeoff. Conversely, with $k=n$, the QRAM is fully parallel, but the qubit count increases significantly, necessitating smaller window sizes ($w'_e, w'_m$). Even with an impractical distance of 3 for algorithmic qubits, some crude estimates show that estimated physical qubit costs would be over 200 million (nearly ten times more than GE's factoring cost). Newer methods of magic state purification might offer cheaper implementations in the future~\cite{gidney2024magic}. We leave this analysis for future work.

\paragraph{Further techniques for reducing resources}
We introduce a technique we call sliced windowing (described in Appendix~\ref{subsec:imp2-slicedwindowing}), which offers two different flavors for performing tradeoff at a logical level by only partially loading a lookup table into memory. 
The first can reduce the number of logical qubits in the lookup register by $50\%$ (which is a $16\%$ reduction in the number of logical qubits). The second one can reduce the number of $\Tof$ gates for addition by $50\%$. While we show that these techniques reduce the number of logical qubits or the $\Tof$ count, they come at the expense of an increased circuit depth. This added depth necessitates higher-distance QEC codes to ensure fault tolerance, which can lead to a net increase in the overall physical resource cost, and hence is not included in our comparison.  Unfortunately, our analysis indicates that sliced windowing is unlikely to reduce the number of \emph{physical} qubits, but we hope that our work can inspire more impactful techniques in the future.

\section{Acknowledgements}
This work started when AS was working at Inveriant Pte.\ Ltd. AS is supported by Innovate UK under grant \emph{10004359}. This work is supported by the National Research Foundation, Singapore, and A*STAR under its Centre for Quantum Technologies Funding Initiative (S24Q2d0009). We also acknowledge funding from the Quantum Engineering Programme (QEP 2.0) under grants \emph{NRF2021-QEP2-02-P05} and \emph{NRF2021-QEP2-02-P01}. We thank Filippo Miatto, Ilan Tzitrin, and Rafael Alexander for useful discussions on resource estimations, Miklos Santha and Michele Orr\`u for useful discussions quantum and classical arithmetic.

\printbibliography

@article{gidney2021factor,
  title={How to factor 2048 bit RSA integers in 8 hours using 20 million noisy qubits},
  author={Gidney, Craig and Eker{\aa}, Martin},
  journal={Quantum},
  volume={5},
  pages={433},
  year={2021},
  publisher={Verein zur F{\"o}rderung des Open Access Publizierens in den Quantenwissenschaften}
}

@article{pham20132d,
  title={A 2D NEAREST-NEIGHBOR QUANTUM ARCHITECTURE FOR FACTORING IN POLYLOGARITHMIC DEPTH},
  author={PHAM, PAUL},
  journal={Quantum Information and Computation},
  volume={13},
  number={11\&12},
  pages={0937--0962},
  year={2013}
}

@article{zalka2006shor,
  title={Shor's algorithm with fewer (pure) qubits},
  author={Zalka, Christof},
  journal={arXiv preprint quant-ph/0601097},
  year={2006}
}

@article{gidney2019windowed,
  title={Windowed quantum arithmetic},
  author={Gidney, Craig},
  journal={arXiv preprint arXiv:1905.07682},
  year={2019}
}

@book{knuth2014art,
  title={The Art of Computer Programming: Seminumerical Algorithms, Volume 2},
  author={Knuth, Donald E},
  year={2014},
  publisher={Addison-Wesley Professional}
}

@article{low2024trading,
  title={Trading T gates for dirty qubits in state preparation and unitary synthesis},
  author={Low, Guang Hao and Kliuchnikov, Vadym and Schaeffer, Luke},
  journal={Quantum},
  volume={8},
  pages={1375},
  year={2024},
  publisher={Verein zur F{\"o}rderung des Open Access Publizierens in den Quantenwissenschaften}
}

@article{doriguello2024practicality,
  title={On the practicality of quantum sieving algorithms for the shortest vector problem},
  author={Doriguello, Joao F and Giapitzakis, George and Luongo, Alessandro and Morolia, Aditya},
  journal={arXiv preprint arXiv:2410.13759},
  year={2024}
}

@article{zhu2024unified,
  title={Unified architecture for a quantum lookup table},
  author={Zhu, Shuchen and Sundaram, Aarthi and Low, Guang Hao},
  journal={arXiv preprint arXiv:2406.18030},
  year={2024}
}

@article{allcock2023constant,
  title={Constant-depth circuits for Uniformly Controlled Gates and Boolean functions with application to quantum memory circuits},
  author={Allcock, Jonathan and Bao, Jinge and Doriguello, Jo{\~a}o F and Luongo, Alessandro and Santha, Miklos},
  journal={arXiv preprint arXiv:2308.08539},
  year={2023}
}

@article{yuan2023optimal,
  title={Optimal (controlled) quantum state preparation and improved unitary synthesis by quantum circuits with any number of ancillary qubits},
  author={Yuan, Pei and Zhang, Shengyu},
  journal={Quantum},
  volume={7},
  pages={956},
  year={2023},
  publisher={Verein zur F{\"o}rderung des Open Access Publizierens in den Quantenwissenschaften}
}

@article{jaques2023qram,
  title={Qram: A survey and critique},
  author={Jaques, Samuel and Rattew, Arthur G},
  journal={arXiv preprint arXiv:2305.10310},
  year={2023}
}

@article{kornerup2021tight,
  title={Tight Bounds on the Spooky Pebble Game: Recycling Qubits with Measurements},
  author={Kornerup, Niels and Sadun, Jonathan and Soloveichik, David},
  journal={arXiv preprint arXiv:2110.08973},
  year={2021}
}

@article{cuccaro2004new,
  title={A new quantum ripple-carry addition circuit},
  author={Cuccaro, Steven A and Draper, Thomas G and Kutin, Samuel A and Moulton, David Petrie},
  journal={arXiv preprint quant-ph/0410184},
  year={2004}
}

@article{gidney2018halving,
  title={Halving the cost of quantum addition},
  author={Gidney, Craig},
  journal={Quantum},
  volume={2},
  pages={74},
  year={2018},
  publisher={Verein zur F{\"o}rderung des Open Access Publizierens in den Quantenwissenschaften}
}

@article{litinski2023compute,
  title={How to compute a 256-bit elliptic curve private key with only 50 million Toffoli gates},
  author={Litinski, Daniel},
  journal={arXiv preprint arXiv:2306.08585},
  year={2023}
}

@article{ragavan2024regev,
  title={Regev Factoring Beyond Fibonacci: Optimizing Prefactors},
  author={Ragavan, Seyoon},
  journal={Cryptology ePrint Archive},
  year={2024}
}

@article{draper2000addition,
  title={Addition on a quantum computer},
  author={Draper, Thomas G},
  journal={arXiv preprint quant-ph/0008033},
  year={2000}
}

@article{gidney2019flexible,
  title={Flexible layout of surface code computations using AutoCCZ states},
  author={Gidney, Craig and Fowler, Austin G},
  journal={arXiv preprint arXiv:1905.08916},
  year={2019}
}

@article{gidney2024magic,
  title={Magic state cultivation: growing T states as cheap as CNOT gates},
  author={Gidney, Craig and Shutty, Noah and Jones, Cody},
  journal={arXiv preprint arXiv:2409.17595},
  year={2024}
}

@article{Haner2020,
  doi = {10.48550/ARXIV.2001.09580},
  url = {https://arxiv.org/abs/2001.09580},
  author = {H\"{a}ner,  Thomas and Jaques,  Samuel and Naehrig,  Michael and Roetteler,  Martin and Soeken,  Mathias},
  title = {Improved quantum circuits for elliptic curve discrete logarithms},
  publisher = {arXiv},
  year = {2020}
}

@article{gouzien2023performance,
  title={Performance analysis of a repetition cat code architecture: Computing 256-bit elliptic curve logarithm in 9 hours with 126 133 cat qubits},
  author={Gouzien, {\'E}lie and Ruiz, Diego and Le R{\'e}gent, Francois-Marie and Guillaud, J{\'e}r{\'e}mie and Sangouard, Nicolas},
  journal={Physical Review Letters},
  volume={131},
  number={4},
  pages={040602},
  year={2023},
  publisher={APS}
}

@article{arunachalam2015robustness,
  title={On the robustness of bucket brigade quantum RAM},
  author={Arunachalam, Srinivasan and Gheorghiu, Vlad and Jochym-O’Connor, Tomas and Mosca, Michele and Srinivasan, Priyaa Varshinee},
  journal={New Journal of Physics},
  volume={17},
  number={12},
  pages={123010},
  year={2015},
  publisher={IOP Publishing}
}

@book{nielsen,
  author = {Michael A. Nielsen and Isaac L. Chuang},
  title = {Quantum Computation and Quantum Information: 10th Anniversary Edition},
  publisher = {Cambridge University Press},
  year = {2011},
  isbn = {9781107002173}
}

@article{Shor1994,
  author = {Shor, Peter W.},
  journal = {Proceedings 35th Annual Symposium on Foundations of Computer Science},
  title = {Algorithms for quantum computation: discrete logarithms and factoring},
  year = {1994},
  volume = {},
  number = {},
  pages = {124-134},
  doi = {10.1109/SFCS.1994.365700}
}

@article{Bennett1973,
  author={Bennett, C. H.},
  journal={IBM Journal of R\&D}, 
  title={Logical Reversibility of Computation}, 
  year={1973},
  volume={17},
  number={6},
  pages={525-532},
  doi={10.1147/rd.176.0525}
}

@article{gidney2019asymptotically,
  title={Asymptotically efficient quantum Karatsuba multiplication},
  author={Gidney, Craig},
  journal={arXiv preprint arXiv:1904.07356},
  year={2019}
}

@article{kahanamoku2024fast,
  title={Fast quantum integer multiplication with zero ancillas},
  author={Kahanamoku-Meyer, Gregory D and Yao, Norman Y},
  journal={arXiv preprint arXiv:2403.18006},
  year={2024}
}

@article{may2019quantum,
  title={Quantum period finding is compression robust},
  author={May, Alexander and Schlieper, Lars},
  journal={arXiv preprint arXiv:1905.10074},
  year={2019}
}

@article{griffiths1996semiclassical,
  title={Semiclassical Fourier transform for quantum computation},
  author={Griffiths, Robert B and Niu, Chi-Sheng},
  journal={Physical Review Letters},
  volume={76},
  number={17},
  pages={3228},
  year={1996},
  publisher={APS}
}

@article{rines2018high,
  title={High performance quantum modular multipliers},
  author={Rines, Rich and Chuang, Isaac},
  journal={arXiv preprint arXiv:1801.01081},
  year={2018}
}

@article{chevignard2024reducing,
  title={Reducing the Number of Qubits in Quantum Factoring},
  author={Chevignard, Cl{\'e}mence and Fouque, Pierre-Alain and Schrottenloher, Andr{\'e}},
  journal={Cryptology ePrint Archive},
  year={2024}
}

@article{ragavan2023optimizing,
  title={Optimizing Space in Regev's Factoring Algorithm},
  author={Ragavan, Seyoon and Vaikuntanathan, Vinod},
  journal={arXiv preprint arXiv:2310.00899},
  year={2023}
}

@article{regev2023efficient,
  title={An efficient quantum factoring algorithm},
  author={Regev, Oded},
  journal={arXiv preprint arXiv:2308.06572},
  year={2023}
}

@inproceedings{ekeraa2024extending,
  title={Extending Regev’s factoring algorithm to compute discrete logarithms},
  author={Eker{\aa}, Martin and G{\"a}rtner, Joel},
  booktitle={International Conference on Post-Quantum Cryptography},
  pages={211--242},
  year={2024},
  organization={Springer}
}

@inproceedings{seifert2001using,
  title={Using fewer qubits in Shor’s factorization algorithm via simultaneous diophantine approximation},
  author={Seifert, Jean-Pierre},
  booktitle={Cryptographers’ Track at the RSA Conference},
  pages={319--327},
  year={2001},
  organization={Springer}
}

@article{horsman2012surface,
  title={Surface code quantum computing by lattice surgery},
  author={Horsman, Dominic and Fowler, Austin G and Devitt, Simon and Van Meter, Rodney},
  journal={New Journal of Physics},
  volume={14},
  number={12},
  pages={123011},
  year={2012},
  publisher={IOP Publishing}
}

@article{wang2024comprehensive,
  title={A Comprehensive Study of Quantum Arithmetic Circuits},
  author={Wang, Siyi and Li, Xiufan and Lee, Wei Jie Bryan and Deb, Suman and Lim, Eugene and Chattopadhyay, Anupam},
  journal={arXiv preprint arXiv:2406.03867},
  year={2024}
}

@article{gidney2019approximate,
  title={Approximate encoded permutations and piecewise quantum adders},
  author={Gidney, Craig},
  journal={arXiv preprint arXiv:1905.08488},
  year={2019}
}

@article{babbush2018encoding,
  title={Encoding electronic spectra in quantum circuits with linear T complexity},
  author={Babbush, Ryan and Gidney, Craig and Berry, Dominic W and Wiebe, Nathan and McClean, Jarrod and Paler, Alexandru and Fowler, Austin and Neven, Hartmut},
  journal={Physical Review X},
  volume={8},
  number={4},
  pages={041015},
  year={2018},
  publisher={APS}
}

@article{jones2013low,
  title={Low-overhead constructions for the fault-tolerant Toffoli gate},
  author={Jones, Cody},
  journal={Physical Review A—Atomic, Molecular, and Optical Physics},
  volume={87},
  number={2},
  pages={022328},
  year={2013},
  publisher={APS}
}

@inproceedings{ekeraa2017quantum,
  title={Quantum algorithms for computing short discrete logarithms and factoring RSA integers},
  author={Eker{\aa}, Martin and H{\aa}stad, Johan},
  booktitle={Post-Quantum Cryptography: 8th International Workshop, PQCrypto 2017, Utrecht, The Netherlands, June 26-28, 2017, Proceedings 8},
  pages={347--363},
  year={2017},
  organization={Springer}
}

@article{giovannetti2008architectures,
  title={Architectures for a quantum random access memory},
  author={Giovannetti, Vittorio and Lloyd, Seth and Maccone, Lorenzo},
  journal={Physical Review A—Atomic, Molecular, and Optical Physics},
  volume={78},
  number={5},
  pages={052310},
  year={2008},
  publisher={APS}
}

@article{luongo2024measurement,
  title={Measurement-based uncomputation of quantum circuits for modular arithmetic},
  author={Luongo, Alessandro and Miti, Antonio Michele and Narasimhachar, Varun and Sireesh, Adithya},
  journal={arXiv preprint arXiv:2407.20167},
  year={2024}
}

@phdthesis{hann2021practicality,
  title={Practicality of quantum random access memory},
  author={Hann, Connor T},
  year={2021},
  school={Yale University}
}

@article{litinski2024quantum,
  title={Quantum schoolbook multiplication with fewer Toffoli gates},
  author={Litinski, Daniel},
  journal={arXiv preprint arXiv:2410.00899},
  year={2024}
}

@article{fowler2018low,
  title={Low overhead quantum computation using lattice surgery},
  author={Fowler, Austin G and Gidney, Craig},
  journal={arXiv preprint arXiv:1808.06709},
  year={2018}
}

@misc{zalka_fast_1998,
    title = {Fast versions of {Shor}'s quantum factoring algorithm},
    url = {http://arxiv.org/abs/quant-ph/9806084},
    doi = {10.48550/arXiv.quant-ph/9806084},
    urldate = {2024-02-06},
    publisher = {arXiv},
    author = {Zalka, Christof},
    month = jun,
    year = {1998},
    note = {arXiv:quant-ph/9806084},
}

\appendix

\section{Larger initial lookup for \texorpdfstring{$\mathsf{RSA}$}{RSA}-\texorpdfstring{$2048$}{2048}}\label{apx:larger-initial-lookup}  

For $\mathsf{RSA}$-$2048$ with parameters \( w_e = 5 \), \( w_m = 5 \), and \( n = 2048 \), the cost of windowed modular exponentiation for a single exponent window is approximately:

\[
    n_{\operatorname{tof}} =  2 \cdot \frac{n}{w_m} \left( C_{\mathsf{Lookup}} + C_{\mathsf{ModAdd}} + C_{\mathsf{Unlookup}} \right) \approx 4.25 \text{ million $\Tof$}
\]

where:
 \( n \): number of bits in the modulus (2048 for $\mathsf{RSA}$-$2048$),
\( w_m \): window size for multiplication (5),
\( C_{\mathsf{Lookup}} = 1024 \): cost of a lookup operation,
\( C_{\mathsf{ModAdd}} = 4096 \): cost of a Cuccaro addition over \( n \) bits,
 \( C_{\mathsf{Unlookup}} = 64 \): cost for creating the unary register representation (32) and applying the phase fixup (32). The multiplicative factor of $2$ in the equation account for the number of circuit calls required for computation and uncomputation of windowed multiplication, while \( \frac{n}{w_m} \) is the number of windows needed to cover the multiplication register. Note that clearing the unary register incurs no additional $\Tof$ cost when using the Logical-AND method for creating the unary representation. For the first \( n'_e \) exponent bits, the total cost of exponentiation is given by:  

\[
    n_{\operatorname{tof}} \approx 4.25 \cdot \frac{n'_e}{w_e} \text{ million $\Tof$}
\]
where \( \frac{n'_e}{w_e} \) is the number of windows over the first \( n'_e \) exponent bits. In contrast, performing a direct lookup over the first \( n'_e \) exponent bits would require \( 2^{n'_e} \) $\Tof$.

\begin{figure}[!ht]
    \centering
    \includegraphics[width=0.75\textwidth]{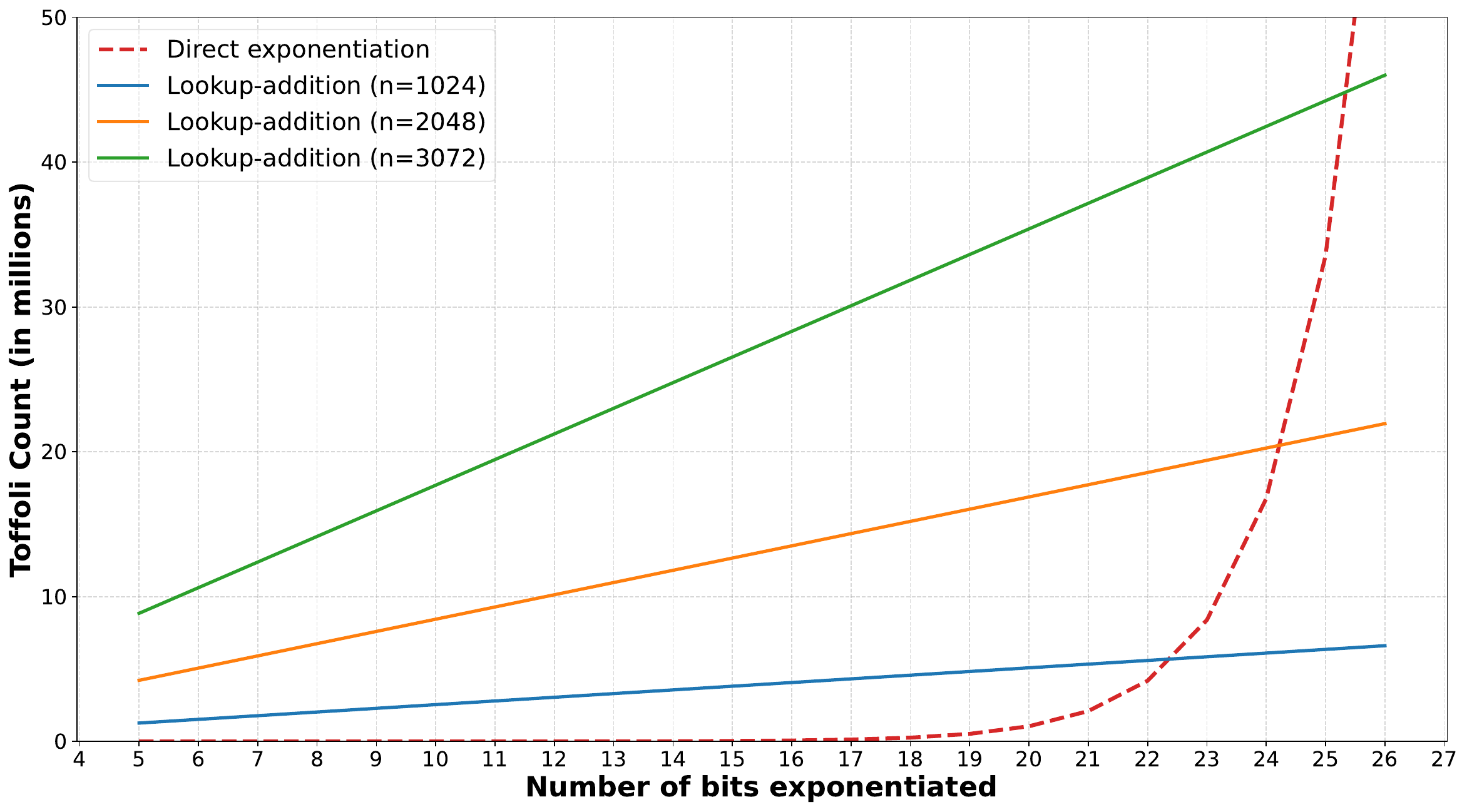}
    \caption{Comparison of the cost of direct exponentiation vs. lookup-addition based exponentiation for the first $n'_e$ bits of the exponent register. The direct exponentiation cost, given by $2^{n'_e}$, is represented by a dashed line, while the lookup-addition based exponentiation cost, given by $2 \cdot \frac{n'_e}{w_e} \cdot \frac{n}{w_m} \cdot (2^{w_e + w_m} + 2n + 2^{(w_e + w_m)/2})$, is shown in solid lines for different values of $n$. Analytical formulas are used to compute the costs.}
    \label{fig:direct-exp-vs-lookup-add-exp}
\end{figure}

\section{Sliced windowing} \label{subsec:imp2-slicedwindowing}

During the windowing stage of GE's factoring algorithm, precomputed lookup table values are loaded into an $n$-qubit lookup register. We propose a method to reduce the number of logical qubits required for these lookups by effectively utilizing carry runways. The GE algorithm employs carry runways to parallelize quantum addition. Assuming the carry runways divide the $n$-qubit quantum register into $m$ pieces, each piece is of size $\frac{n}{m}$. Our approach involves loading only $\frac{n}{m}$ or a multiple of $\frac{n}{m}$ bits of the lookup table at a time. We then add this loaded register of size $\frac{n}{m}$ to the relevant quantum register, uncompute the values, and load the next $\frac{n}{m}$ bits. This strategy reduces the abstract qubit count of the GE algorithm from $3n$ to $2n + \frac{n}{m}$. Although this increases the $\Tof$ count of the lookups by a factor of $m$, it can be managed by rearranging the order of the lookups and additions. We propose two variants of the sliced windowing scheme that reduce the logical qubit count without impacting the $\Tof$ count.

\subsection{Variant A: logical qubit reduction with sliced windowing}

\begin{figure}
    \centering
    \begin{subfigure}[b]{\textwidth}
        \centering
        \includegraphics[width=\linewidth]{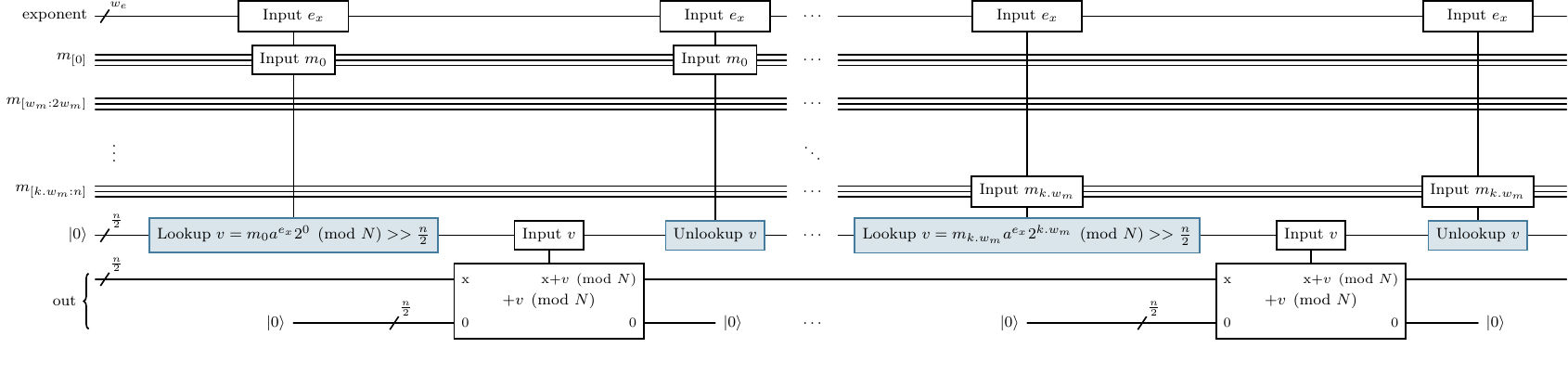}
    \end{subfigure}
    
    \vspace{5pt}
    
    \begin{subfigure}[b]{\textwidth}
        \centering
        \includegraphics[width=\linewidth]{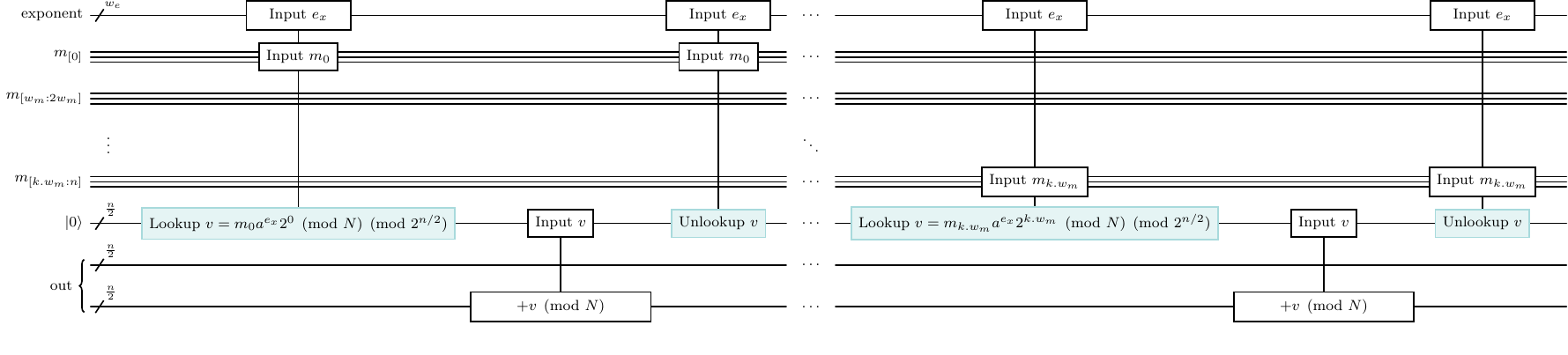}
    \end{subfigure}
    
    \caption{Variant A - The two steps of the sliced windowing improvement. In the first part (upper figure, in dark blue), we use an adder (e.g., Gidney's adder) that reduces the $\Tof$ cost at the expense of using more space. In the lower figure (in light blue), we perform additions in the second part of the target register using an adder circuit that utilizes no ancilla qubits.}
    \label{fig:sliced-windowing-both}
\end{figure}

In the original windowing circuit for a number $N$ (with $n = \log N$), let the exponent register $e$ have a window size $w_e = \lfloor \frac{1}{2}\log n \rfloor$, a the multiplication register $m$ (of size $n$) have a window size $w_m = \lfloor \frac{1}{2}\log n \rfloor$. Assume we use a lookup register $\ell$ of size $n/2$ (instead of the general size $n$), and a target register $t$ of size $n$ for adding the lookup values. In this variant, we perform lookup additions on one half of the target register $t[0:n/2]$, using the second half, $t[n/2:n]$, as ancillas required for Gidney's logical-AND-based addition~\cite{gidney2018halving}. Lookups can be conducted by windowing over the multiplication register $m$, with a memory size of $n/2$, compared to the original $n$ size proposed in~\cite{gidney2019windowed}. The lookup costs $2^{w_e + w_m} = 2^{2\lfloor \frac{1}{2}\log n \rfloor}$ in $\Tof$ gates and depth. The looked-up values can then be added to the target register using carry runways that attach register pieces of size $n/4$. The other half of the target register, of size $n/2$, serves as the ancillas needed for Gidney's logical-AND-based addition. A sweep over the entire multiplication register is performed while only looking up half the memory bits of the lookup table. The $\Tof$ count for the Gidney additions of registers of size $x$ is $x$. In our case, this addition costs $n/2$, with a measurement depth of $n$. In the second phase of the lookup additions, we perform another set of lookups, which again cost $2^{w_e + w_m} = 2^{2\lfloor \frac{1}{2}\log n \rfloor}$ in $\Tof$ gates and depth, with a size of $n/2$ (shown in green in Figure~\ref{fig:sliced-windowing-both}). Here, we use a normal Cuccaro addition (with no ancillas). The $\Tof$ count for the Cuccaro RCA additions of registers of size $x$ is $2x$, with a measurement depth of $2x$. Thus, the additions cost $2 \cdot n/2 = n$ $\Tof$ gates with a measurement depth of $2 \cdot n/2 = n$. Overall, this protocol achieves \emph{no reduction} in the $\Tof$ count but results in a \emph{decrease} of $n/2$ logical qubits and an \emph{increase} in depth by $n$.

\subsection{Variant B: reintroducing temporarily protected ancillas for a lower toffoli count}

In this variant, we aim to lower the $\Tof$ count by strategically reusing ancillas that are idle for part of the computation. Specifically, we repurpose ancillas saved in Variant A by only loading half of the lookup table at a time, and then reintroduce them during the addition to the second half of the target register. In Variant A, when adding the first half of the lookup values to the target register, the second half of the target register served as ancillas for logical-$\mathsf{AND}$ adder~\cite{gidney2018halving}. When processing the second half, we had to switch to Cuccaro's ripple-carry adder because no ancillas were left available. In this variant, we avoid switching to Cuccaro's adder by reintroducing the ancillas that were saved during the first stage of the computation. These ancillas, which would otherwise remain idle for the first half of the algorithm's runtime, are now only needed and protected during the second half, when they are essential for performing the logical-$\mathsf{AND}$ addition into the second half of the target register. This results in a reduction in the $\Tof$ count by $n/2$ (per lookup-addition), while maintaining the same logical qubit count. The logical qubits need to effectively be protected for only half of the runtime and are ``borrowed'' for the second half of the process, where they are used to perform the addition on the second half of the target register. The tradeoff is that the circuit depth increases (so it is possible that we end up having to protect the logical information for longer, thus defeating the purpose of the idle ancillas), depending on the window size, but we gain an overall reduction in $\Tof$ cost compared to the baseline Cuccaro adder approach.
\begin{figure}
    \centering
    \includegraphics[width=0.5\textwidth]{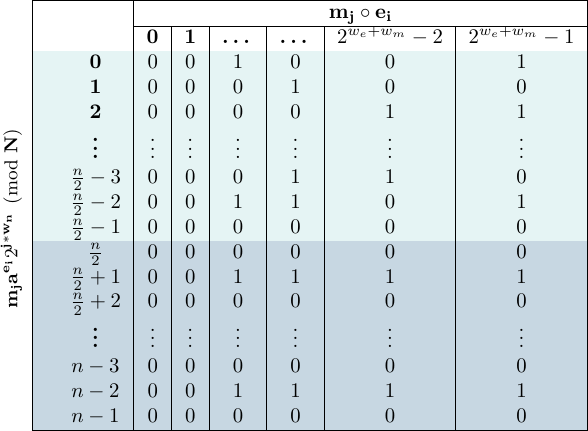}
    \vspace{0.2cm}
    \caption{Table of lookup values for sliced windowing in the quantum modular exponentiation circuit with modulus $N$ (of size $n=2048$ bits) by GE. Here, $w_e$ and $w_m$ represent the window sizes of the exponent and multiplication registers. The value stored in each column is of the form $m_j a^{e_i2^{i\cdot w_e}} 2^{j\cdot w_m} \pmod N$}
    \label{fig:sliced-windowing-table}
\end{figure}
\newpage
\section{Analysing tradeoffs on physical resources}\label{apx:mqb_tradeoffs_ge_v_ours}

\begin{table}[htbp]
\centering
\begin{adjustbox}{width=\textwidth}
\begin{tabular}{c|c|c|c|c|c|c|c|c||c|c|c|c|c|c|c}
$\bm{n}$ & $\bm{n_e}$ & \textbf{gate err} & $L_1$ & $L_2$ & $d_{\mathrm{off}}$ & $g_{\mathrm{mul}}$ & $g_{\mathrm{exp}}$ & $g_{\mathrm{sep}}$ & \% & v.p.r & $\mathbb{E}[\text{vol}]$ & Mqb & hrs & $\mathbb{E}[\text{hrs}]$ & B Tofs\\
\hline
$2048$    &$3029$    &$10^{-3}$   &$15$      &$27$      &$2$       &$5$       &$5$       &$2048$    &$36\%$    &$4.861$   &$7.596$   &$14.747$  &$7.911$   &$12.361$  &$2.656$   \\
$2048$    &$3029$    &$10^{-3}$   &$17$      &$27$      &$6$       &$5$       &$5$       &$2048$    &$25\%$    &$5.157$   &$6.876$   &$15.592$  &$7.938$   &$10.584$  &$2.665$   \\
$2048$    &$3029$    &$10^{-3}$   &$15$      &$29$      &$2$       &$5$       &$5$       &$2048$    &$18\%$    &$5.878$   &$7.168$   &$17.492$  &$8.064$   &$9.834$   &$2.656$   \\
$2048$    &$3029$    &$10^{-3}$   &$17$      &$29$      &$6$       &$5$       &$5$       &$2048$    &$4\%$     &$6.241$   &$6.501$   &$18.513$  &$8.091$   &$8.428$   &$2.665$   \\
$2048$    &$3029$    &$10^{-3}$   &$15$      &$27$      &$4$       &$5$       &$5$       &$1024$    &$31\%$    &$4.047$   &$5.865$   &$19.249$  &$5.046$   &$7.313$   &$2.698$   \\
$2048$    &$3029$    &$10^{-3}$   &$17$      &$27$      &$4$       &$5$       &$5$       &$1024$    &$21\%$    &$4.545$   &$5.753$   &$21.616$  &$5.046$   &$6.388$   &$2.698$   \\
$2048$    &$3029$    &$10^{-3}$   &$15$      &$29$      &$3$       &$5$       &$5$       &$1024$    &$18\%$    &$5.194$   &$6.334$   &$24.001$  &$5.194$   &$6.334$   &$2.695$   \\
$2048$    &$3029$    &$10^{-3}$   &$17$      &$29$      &$5$       &$4$       &$5$       &$1024$    &$4\%$     &$5.424$   &$5.65$    &$25.265$  &$5.153$   &$5.367$   &$3.041$   \\
$2048$    &$3029$    &$10^{-3}$   &$17$      &$31$      &$6$       &$4$       &$5$       &$1024$    &$2\%$     &$5.981$   &$6.103$   &$27.308$  &$5.256$   &$5.364$   &$3.045$   \\
$2048$    &$3029$    &$10^{-3}$   &$15$      &$27$      &$3$       &$4$       &$5$       &$512$     &$32\%$    &$3.946$   &$5.802$   &$29.184$  &$3.245$   &$4.772$   &$3.125$   \\
$2048$    &$3029$    &$10^{-3}$   &$17$      &$27$      &$4$       &$4$       &$5$       &$512$     &$19\%$    &$4.415$   &$5.45$    &$32.602$  &$3.25$    &$4.012$   &$3.132$   \\
$2048$    &$3029$    &$10^{-3}$   &$17$      &$29$      &$6$       &$4$       &$4$       &$512$     &$4\%$     &$5.576$   &$5.808$   &$40.075$  &$3.339$   &$3.478$   &$3.718$   \\
\end{tabular}
\end{adjustbox}
\begin{adjustbox}{width=\textwidth}
\begin{tabular}{c|c|c|c|c|c|c|c|c||c|c|c|c|c|c|c}
$\bm{n}$ & $\bm{n_e}$ & \textbf{gate err} & $L_1$ & $L_2$ & $d_{\mathrm{off}}$ & $g_{\mathrm{mul}}$ & $g_{\mathrm{exp}}$ & $g_{\mathrm{sep}}$ & \% & v.p.r & $\mathbb{E}[\text{vol}]$ & Mqb & hrs & $\mathbb{E}[\text{hrs}]$ & B Tofs \\
\hline
$2048$    &$3029$    &$10^{-3}$   &$15$      &$27$      &$2$       &$5$       &$5$       &$2048$    &$35\%$    &$4.757$   &$7.319$   &$14.747$  &$7.743$   &$11.912$  &$2.607$   \\
$2048$    &$3029$    &$10^{-3}$   &$17$      &$27$      &$3$       &$5$       &$5$       &$2048$    &$25\%$    &$5.034$   &$6.712$   &$15.592$  &$7.749$   &$10.332$  &$2.609$   \\
$2048$    &$3029$    &$10^{-3}$   &$15$      &$29$      &$8$       &$5$       &$5$       &$2048$    &$17\%$    &$5.776$   &$6.959$   &$17.492$  &$7.925$   &$9.548$   &$2.621$   \\
$2048$    &$3029$    &$10^{-3}$   &$17$      &$29$      &$6$       &$5$       &$5$       &$2048$    &$4\%$     &$6.103$   &$6.357$   &$18.513$  &$7.912$   &$8.242$   &$2.616$   \\
$2048$    &$3029$    &$10^{-3}$   &$15$      &$27$      &$2$       &$5$       &$5$       &$1024$    &$31\%$    &$3.918$   &$5.679$   &$19.249$  &$4.886$   &$7.08$    &$2.643$   \\
$2048$    &$3029$    &$10^{-3}$   &$17$      &$27$      &$6$       &$5$       &$5$       &$1024$    &$20\%$    &$4.419$   &$5.524$   &$21.616$  &$4.906$   &$6.133$   &$2.656$   \\
$2048$    &$3029$    &$10^{-3}$   &$15$      &$29$      &$6$       &$5$       &$5$       &$1024$    &$17\%$    &$5.05$    &$6.084$   &$24.001$  &$5.049$   &$6.084$   &$2.656$   \\
$2048$    &$3029$    &$10^{-3}$   &$17$      &$29$      &$5$       &$4$       &$5$       &$1024$    &$4\%$     &$5.247$   &$5.466$   &$25.265$  &$4.984$   &$5.192$   &$2.988$   \\
$2048$    &$3029$    &$10^{-3}$   &$17$      &$31$      &$6$       &$4$       &$5$       &$1024$    &$2\%$     &$5.778$   &$5.895$   &$27.308$  &$5.078$   &$5.181$   &$2.992$   \\
$2048$    &$3029$    &$10^{-3}$   &$15$      &$27$      &$4$       &$4$       &$5$       &$512$     &$31\%$    &$3.773$   &$5.468$   &$29.184$  &$3.103$   &$4.497$   &$3.079$   \\
$2048$    &$3029$    &$10^{-3}$   &$17$      &$27$      &$5$       &$4$       &$5$       &$512$     &$18\%$    &$4.221$   &$5.148$   &$32.602$  &$3.108$   &$3.79$    &$3.085$   \\
$2048$    &$3029$    &$10^{-3}$   &$17$      &$29$      &$5$       &$4$       &$5$       &$512$     &$4\%$     &$5.336$   &$5.558$   &$40.075$  &$3.196$   &$3.329$   &$3.085$   \\
\end{tabular}
\end{adjustbox}
\caption{Table with increasing qubit counts for GE (top) vs.\ GE incorporating our improvements (bottom). Here, we explore the lowest expected runtimes that can be achieved for a given number of physical qubits (\textbf{gate err} $10^{-3}$).}\label{tab:impact_qecc_superconducting_increasing_qubit_count}\end{table}

\begin{table}[htbp]
\centering
\begin{adjustbox}{width=\textwidth}
\begin{tabular}{c|c|c|c|c|c|c|c|c||c|c|c|c|c|c|c}
$\bm{n}$ & $\bm{n_e}$ & \textbf{gate err} & $L_1$ & $L_2$ & $d_{\mathrm{off}}$ & $g_{\mathrm{mul}}$ & $g_{\mathrm{exp}}$ & $g_{\mathrm{sep}}$ & \% & v.p.r & $\mathbb{E}[\text{vol}]$ & Mqb & hrs & $\mathbb{E}[\text{hrs}]$ & B Tofs\\
\hline
$2048$    &$3029$    &$10^{-4}$  &$7$       &$13$      &$3$       &$5$       &$6$       &$2048$    &$27\%$    &$0.866$   &$1.187$   &$3.195$   &$6.507$   &$8.914$   &$2.658$   \\
$2048$    &$3029$    &$10^{-4}$  &$9$       &$13$      &$3$       &$5$       &$6$       &$2048$    &$20\%$    &$0.99$    &$1.238$   &$3.651$   &$6.507$   &$8.134$   &$2.658$   \\
$2048$    &$3029$    &$10^{-4}$  &$7$       &$13$      &$3$       &$5$       &$5$       &$1024$    &$23\%$    &$0.658$   &$0.855$   &$3.977$   &$3.972$   &$5.159$   &$2.695$   \\
$2048$    &$3029$    &$10^{-4}$  &$9$       &$13$      &$3$       &$5$       &$5$       &$1024$    &$15\%$    &$0.779$   &$0.916$   &$4.706$   &$3.972$   &$4.673$   &$2.695$   \\
$2048$    &$3029$    &$10^{-4}$  &$7$       &$13$      &$4$       &$5$       &$5$       &$512$     &$21\%$    &$0.558$   &$0.706$   &$5.273$   &$2.538$   &$3.212$   &$2.774$   \\
$2048$    &$3029$    &$10^{-4}$  &$7$       &$15$      &$5$       &$5$       &$5$       &$512$     &$12\%$    &$0.731$   &$0.831$   &$6.513$   &$2.694$   &$3.062$   &$2.78$    \\
$2048$    &$3029$    &$10^{-4}$  &$9$       &$13$      &$5$       &$5$       &$5$       &$512$     &$13\%$    &$0.756$   &$0.869$   &$7.141$   &$2.542$   &$2.922$   &$2.78$    \\
$2048$    &$3029$    &$10^{-4}$  &$9$       &$15$      &$5$       &$4$       &$5$       &$512$     &$2\%$     &$0.847$   &$0.865$   &$7.621$   &$2.668$   &$2.723$   &$3.139$   \\
\end{tabular}
\end{adjustbox}
\begin{adjustbox}{width=\textwidth}
\begin{tabular}{c|c|c|c|c|c|c|c|c||c|c|c|c|c|c|c}
$\bm{n}$ & $\bm{n_e}$ & \textbf{gate err} & $L_1$ & $L_2$ & $d_{\mathrm{off}}$ & $g_{\mathrm{mul}}$ & $g_{\mathrm{exp}}$ & $g_{\mathrm{sep}}$ & \% & v.p.r & $\mathbb{E}[\text{vol}]$ & Mqb & hrs & $\mathbb{E}[\text{hrs}]$ & B Tofs \\
\hline
$2048$    &$3029$    &$10^{-4}$  &$7$       &$13$      &$2$       &$5$       &$6$       &$2048$    &$27\%$    &$0.849$   &$1.163$   &$3.195$   &$6.378$   &$8.737$   &$2.594$   \\
$2048$    &$3029$    &$10^{-4}$  &$9$       &$13$      &$2$       &$5$       &$6$       &$2048$    &$20\%$    &$0.97$    &$1.213$   &$3.651$   &$6.378$   &$7.972$   &$2.594$   \\
$2048$    &$3029$    &$10^{-4}$  &$7$       &$13$      &$9$       &$5$       &$5$       &$1024$    &$22\%$    &$0.649$   &$0.832$   &$3.977$   &$3.918$   &$5.023$   &$2.666$   \\
$2048$    &$3029$    &$10^{-4}$  &$9$       &$13$      &$6$       &$5$       &$5$       &$1024$    &$15\%$    &$0.766$   &$0.89$    &$4.706$   &$3.904$   &$4.539$   &$2.656$   \\
$2048$    &$3029$    &$10^{-4}$  &$7$       &$13$      &$3$       &$5$       &$5$       &$512$     &$21\%$    &$0.541$   &$0.684$   &$5.273$   &$2.461$   &$3.115$   &$2.72$    \\
$2048$    &$3029$    &$10^{-4}$  &$7$       &$15$      &$4$       &$5$       &$5$       &$512$     &$12\%$    &$0.708$   &$0.804$   &$6.513$   &$2.608$   &$2.963$   &$2.725$   \\
$2048$    &$3029$    &$10^{-4}$  &$9$       &$13$      &$4$       &$5$       &$5$       &$512$     &$13\%$    &$0.733$   &$0.843$   &$7.141$   &$2.465$   &$2.833$   &$2.725$   \\
$2048$    &$3029$    &$10^{-4}$  &$9$       &$15$      &$5$       &$4$       &$5$       &$512$     &$2\%$     &$0.82$    &$0.836$   &$7.621$   &$2.581$   &$2.634$   &$3.085$   \\

\end{tabular}
\end{adjustbox}
\caption{Table with increasing qubit counts for our algorithm for GE (top) vs.\ GE incorporating our improvements (bottom). Here, we explore the lowest expected runtimes that can be achieved for a given number of physical qubits (\textbf{gate err} $10^{-4}$).}\label{tab:impact_qecc_superconducting_increasing_qubit_count10-4err}\end{table}

\begin{table}[ht] \centering \begin{adjustbox}{width=0.85\textwidth} \begin{tabular}{c|c|c|c|c|c|c|c|c|c} \hline $\bm{n}$ & $\bm{n}_e$ & \textbf{gate err} & Mqb & \multicolumn{2}{c|}{$\mathbb{E}[\text{hrs}]$} & \% improvement & \multicolumn{2}{c|}{B Tofs} & \% improvement \\ \cline{5-6} \cline{8-9} & & & & GE & Ours & $\mathbb{E}[\text{hrs}]$ & GE & Ours & in B Tofs \\ \hline $2048$ & $3029$ & $10^{-4}$ & $3.195$ & $8.914$ & $8.737$ & $1.99\%$ & $2.658$ & $2.594$ & $2.41\%$ \\ $2048$ & $3029$ & $10^{-4}$ & $3.651$ & $8.134$ & $7.972$ & $1.99\%$ & $2.658$ & $2.594$ & $2.41\%$ \\ $2048$ & $3029$ & $10^{-4}$ & $3.977$ & $5.159$ & $5.023$ & $2.64\%$ & $2.695$ & $2.666$ & $1.08\%$ \\ $2048$ & $3029$ & $10^{-4}$ & $4.706$ & $4.673$ & $4.539$ & $2.87\%$ & $2.695$ & $2.656$ & $1.45\%$ \\ $2048$ & $3029$ & $10^{-4}$ & $5.273$ & $3.212$ & $3.115$ & $3.02\%$ & $2.774$ & $2.720$ & $1.95\%$ \\ $2048$ & $3029$ & $10^{-4}$ & $6.513$ & $3.062$ & $2.963$ & $3.23\%$ & $2.780$ & $2.725$ & $1.98\%$ \\ $2048$ & $3029$ & $10^{-4}$ & $7.141$ & $2.922$ & $2.833$ & $3.05\%$ & $2.780$ & $2.725$ & $1.98\%$ \\ $2048$ & $3029$ & $10^{-4}$ & $7.621$ & $2.723$ & $2.634$ & $3.27\%$ & $3.139$ & $3.085$ & $1.72\%$ \\ \hline \end{tabular} 
\end{adjustbox}
\caption{Comparison in $\%$ of the improvements of Table~\ref{tab:impact_qecc_superconducting_increasing_qubit_count10-4err} (\textbf{gate err} $10^{-4}$). Every row in this table compares the improvements for the expected runtime and $\Tof$ count (in billions)  for a fixed number of physical qubits (Mqb).}
\label{tab:comparing_runtime_and_btofs_improvements10-4err}
\end{table}
\begin{figure}[ht]
    \centering
    \includegraphics[width=0.9\textwidth]{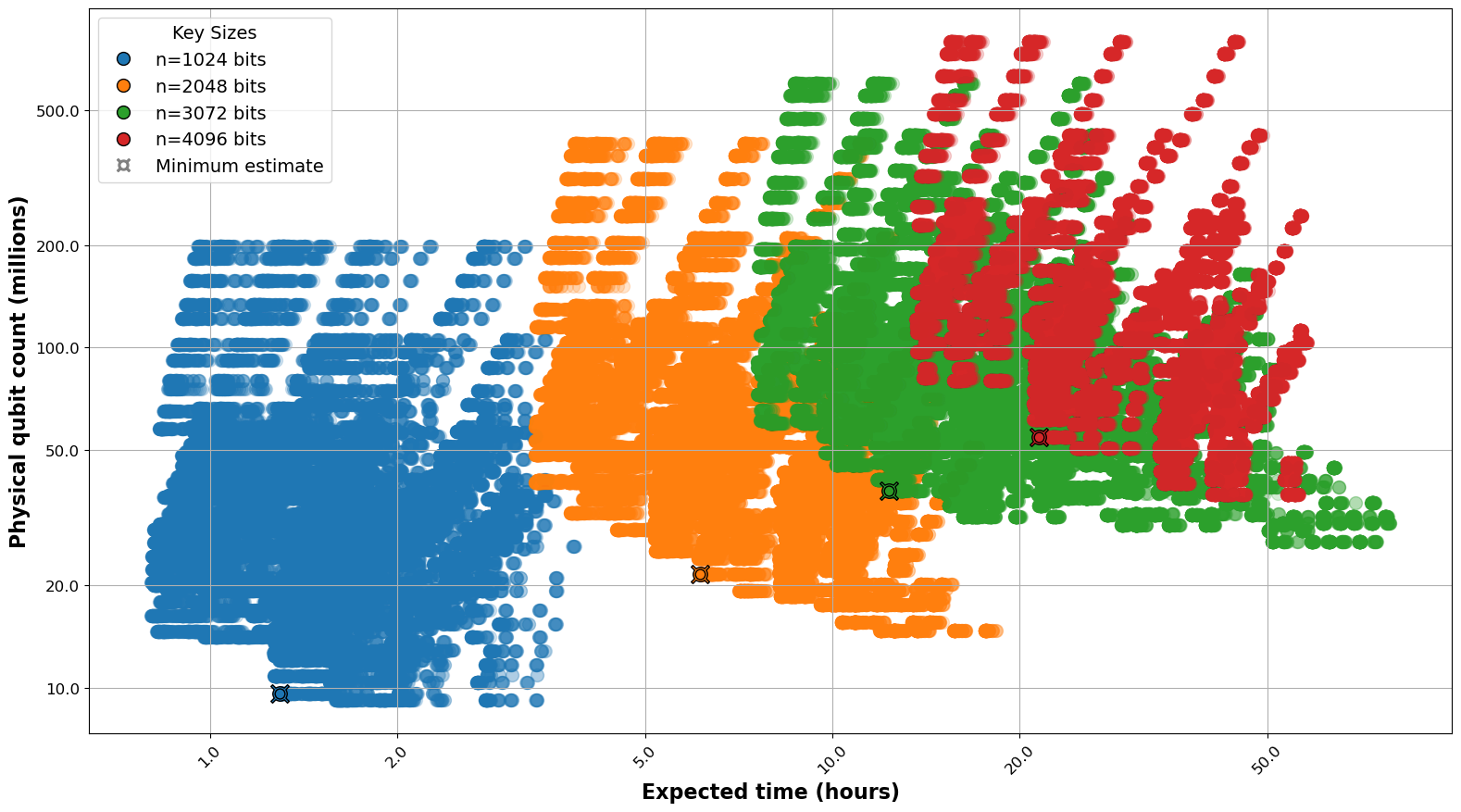}
    \caption{A view of all possible tradeoffs for expected runtime and physical qubit count to break $\mathsf{RSA}$ of various key sizes when using our optimizations in GE factoring algorithm. Here, the physical error rate is set to $10^{-3}$.}
    \label{fig:all-surviving-estimates}
\end{figure}

\newpage
\section{Pseudocode of useful subroutines}\label{apx:pseudocode}

\begin{table}[ht]
  \centering
  
\begin{algorithm}[H]
\caption{Standard Multicontrolled Lookup}
\label{alg:standard_lookup}
\SetKwFunction{STANDARDLOOKUP}{STANDARD\_LOOKUP}
\KwIn{Address register $\underline{a}$, classical lookup table $\operatorname{table}$, lookup register $\underline{l}$}
\KwOut{Updated lookup register $\underline{l}$ with values from $\operatorname{table}$ corresponding to $\underline{a}$}
\SetKwProg{Fn}{Function}{:}{}
\Fn{\STANDARDLOOKUP{$\underline{a}$, table, $\underline{l}$}}{
    \For{$i \leftarrow 0$ \KwTo $2^{|\underline{a}|} - 1$}{
        \If{$\underline{a}$ is in state $|i\rangle$}{
            $\underline{l} \leftarrow \operatorname{table}[i]$
        }
    }
}
\end{algorithm}

  \caption{Lookup table pseudocode}
  \label{lst:pseudocode_lookuptable}
\end{table}
\begin{table}[p]
  \centering
  \begin{algorithm}[H]
\caption{Unlookup Unitary}
\label{alg:unlookup_unitary}

\SetKwFunction{INITUNARY}{INIT\_UNARY}
\SetKwFunction{LOOKUP}{LOOKUP}
\SetKwFunction{UNLOOKUP}{UNLOOKUP}
\KwIn{Address register $\underline{a}$, classical lookup table $\operatorname{table}$, lookup register $\underline{l}$}
\KwOut{Cleared lookup register $\underline{l}$ with applied phase corrections}

\SetKwProg{Fn}{Function}{:}{}
\Fn{\INITUNARY{$\underline{b}$, $\underline{c}$}}{
    \tcp{Initialize an out-of-place unary mapping from $\underline{b}$ to $\underline{c}$}
    \uIf{$\underline{c} \neq 0^{\otimes 2^{|\underline{b}|}}$}{
        \textbf{assertion failed}
    }
    $\underline{c_0} \leftarrow 1$\;
    \For{$i \leftarrow 0$ \KwTo $|\underline{b}| - 1$}{
        \For{$j \leftarrow 0$ \KwTo $2^i - 1$}{
            \If{$\underline{b_i}$}{
                swap($\underline{c_j}, \underline{c_{j+2^i}}$)\;
            }
        }
    }
}

\Fn{\UNLOOKUP{$\underline{a}$, table, $\underline{l}$}}{
    \tcp{Clear the lookup register $\underline{l}$ and apply phase corrections}
    $\mathcal{H}(\underline{l})$\;
    $l \leftarrow \underline{l}$\;
    $\underline{l} \leftarrow 0^{\otimes 2^{|\underline{l}|}}$\;
    \INITUNARY{$\underline{a}[|\underline{a}|/2:], \underline{l}[0:2^{|\underline{a}|/2}]$}\;
    phases $\leftarrow \operatorname{table}^T \cdot l$\;
    $F \leftarrow \text{array.zeros}(2^{|\underline{a}|/2}, 2^{|\underline{a}|/2})$\;
    \For{$\operatorname{addr}_t \leftarrow 0$ \KwTo $\operatorname{len}(\operatorname{table}) - 1$}{
        \If{phases[$t$] $= -1$}{
            $F[\operatorname{addr}_t[|\underline{a}|/2:], \operatorname{addr}_t[0:|\underline{a}|/2]] \leftarrow 1$\;
        }
    }
    $\mathcal{H}(\underline{l}[0:2^{|\underline{a}|/2}])$\;
    \LOOKUP{$\underline{a}[0:|\underline{a}|/2], F, \underline{l}[0:2^{|\underline{a}|/2}]$}\;
    $\mathcal{H}(\underline{l}[0:2^{|\underline{a}|/2}])$\;
    reverse \INITUNARY{$\underline{a}[|\underline{a}|/2:], \underline{l}[0:2^{|\underline{a}|/2}]$}\;
}
\end{algorithm}

  \caption{Unlookup pseudocode}
  \label{lst:unlookup}
\end{table}

\begin{table}[p]
  \centering
  \begin{algorithm}[H]
\caption{Windowed Quantum Modular Exponentiation}
\label{alg:windowed-mod-exp}

\SetKwFunction{MODULARMULTINVERSE}{MODULAR\_MULT\_INVERSE}
\SetKwFunction{LOOKUPTABLE}{LOOKUP\_TABLE}
\SetKwFunction{QROMLOOKUP}{QROM\_LOOKUP}
\SetKwFunction{INPLACEADDMOD}{INPLACE\_ADD\_MOD}
\SetKwFunction{QRAMLOOKUP}{QRAM\_LOOKUP}
\SetKwFunction{UNLOOKUP}{UNLOOKUP}
\SetKwFunction{POP}{POP}
\SetKwFunction{WINDOWEDMODULAREXPONENTIATION}{WINDOWED\_MODULAR\_EXPONENTIATION}

\KwIn{Modulus $N$, base $g$, target register $\underline{t}$, exponent register $\underline{e}$, multiplication register $\underline{m}$, lookup register $\underline{l}$, exponent window size $w_e$, multiplication window size $w_m$}
\KwOut{Final exponentiated state $\ket{e} \rightarrow \ket{e}\ket{g^e \pmod{N}}$}

\SetKwProg{Fn}{Function}{:}{}
\Fn{\WINDOWEDMODULAREXPONENTIATION{$N$, $g$, $\underline{t}$, $\underline{e}$, $\underline{m}$, $\underline{l}$, $w_e$, $w_m$}}{
    \tcp{Compute modular inverse of $g$ modulo $N$}
    $g_i \leftarrow \MODULARMULTINVERSE(g, N)$\;
    \textbf{assert} $g_i \neq \text{None}$\;
    
    \For{$x \leftarrow 0$ \KwTo $\lceil \frac{\text{len}(\underline{e})}{w_e} \rceil - 1$}{
        \tcp{Create reference to the exponent window}
        $\underline{\text{e}}_{\text{window}} \leftarrow \underline{\text{e}}.\POP(w_e)$\;
        
        \For{$y \leftarrow 0$ \KwTo $\lceil \frac{\text{len}(\underline{m})}{w_m} \rceil - 1$}{
            \tcp{Create reference to the multiplication window}
            $\underline{m}_{\text{window}} \leftarrow \underline{m}.\POP(w_m)$\;
            $table \leftarrow \LOOKUPTABLE(g, x, w_e, y, w_m)$\;

            \tcp{Perform lookup addition}
            \QROMLOOKUP{$\underline{m}_{\text{window}} \parallel \underline{e}_{\text{window}}, table, \underline{l}$}\;
            \INPLACEADDMOD{$\underline{t}, \underline{l}$}\;
            \UNLOOKUP{$\underline{m}_{\text{window}} \parallel \underline{e}_{\text{window}}, table, \underline{l}$}\;
        }
        \tcp{Swap target and multiplication registers}
            $\underline{t}, \underline{m} \leftarrow \underline{m}, \underline{t}$\;

        \tcp{Uncompute register $m$}
        \For{$y \leftarrow 0$ \KwTo $\lceil \frac{\text{len}(\underline{m})}{w_m} \rceil - 1$}{
            
            $\underline{m}_{\text{window}} \leftarrow \underline{m}.\POP(w_m)$\;
            $table \leftarrow \LOOKUPTABLE(g_i, x, w_e, y, w_m)$\;

            \tcp{Perform lookup subtraction}
            \QRAMLOOKUP{$\underline{m}_{\text{window}} \parallel \underline{e}_{\text{window}}, table, \underline{l}$}\;
            reverse \INPLACEADDMOD{$\underline{t}, \underline{l}$}\;
            \UNLOOKUP{$\underline{m}_{\text{window}} \parallel \underline{e}_{\text{window}}, table, \underline{l}$}\;
        }
    }
}
\end{algorithm}

  \caption{Windowed modular exponentiation pseudocode}
  \label{lst:windowed_quantum_modular_exponentiation}
\end{table}

\begin{table}[p]
  \centering
\begin{algorithm}[H]
\caption{Phase unlookup}
\label{alg:phase_unlookup}
\SetKwFunction{CPUT}{COMPUTE\_PHASE\_UNLOOKUP\_TABLE}
\caption{Phase Unlookup Table}
\label{alg:phase_unlookup_table}
\KwIn{window size of exponent $w_e$, window size of multiplication register $w_m$, the table we used for the lookup operation $table$, the quantum register holding the lookup value $\underline{l}$}
\KwOut{a classical table containing information phase corrections $F$}
\SetKwProg{Fn}{Function}{:}{}
\Fn{\CPUT{$\underline{a}$, table, $\underline{l}$}}{
\texttt{// Measure lookup register in X basis} 
$\mathcal{H}(\underline{l})$ \; 
$l \leftarrow \underline{l}$ \;
$\underline{l} \leftarrow 0^{\otimes 2^{\text{len}(\underline{l})}}$ \;
 \texttt{// Compute phases from measurement}
phases $\leftarrow \text{table}^T.l$ \;

$F \leftarrow \text{array.zeros}(2^{w_e}, 2^{w_m})$ \; \texttt{// Init phase correction table}

\For{$addr_t \leftarrow 0$ \KwTo $len(table) - 1$}{
    \If{$\text{phases}[addr_t] == -1$}{
        $e_{\text{val}} \leftarrow addr_t[0:w_e]$ \;
        $m_{\text{val}} \leftarrow addr_t[w_e:w_e + w_m]$ \;
        $F[e_{\text{val}}, m_{\text{val}}] \leftarrow 1$ \;
    }
}
\Return{$F$} \;}
\end{algorithm}

\begin{algorithm}[H]
\caption{Windowed Quantum Modular Exponentiation with Deferred Uncomputation}
\label{alg:windowed_quantum_modular_exponentiation}

\SetKwFunction{MODULARMULTINVERSE}{MODULAR\_MULT\_INVERSE}
\SetKwFunction{LOOKUPTABLE}{LOOKUP\_TABLE}
\SetKwFunction{QRAMLOOKUP}{QRAM\_LOOKUP}
\SetKwFunction{INPLACEADDMOD}{INPLACE\_ADD\_MOD}
\SetKwFunction{CPUT}{COMPUTE\_PHASE\_UNLOOKUP\_TABLE}
\SetKwFunction{INITUNARY}{INIT\_UNARY}
\SetKwFunction{LOOKUP}{LOOKUP}
\SetKwFunction{POP}{POP}
\SetKwFunction{DWME}{DU\_WINDOWED\_MODULAR\_EXPONENTIATION}
\KwIn{Modulus $N$, base $g$, target register $\underline{t}$, exponent register $\underline{e}$, multiplication register $\underline{m}$, lookup register $\underline{l}$, exponent window size $w_e$, multiplication window size $w_m$}
\KwOut{$F$}
\SetKwProg{Fn}{Function}{:}{}
\Fn{\DWME{$N$, $g$, $\underline{t}$, $\underline{e}$, $\underline{m}$, $\underline{l}$, $w_e$, $w_m$}}{
    $g_i \leftarrow \MODULARMULTINVERSE(g, N)$ \;
    \textbf{assert} $g_i \neq \texttt{None}$
    
    $F \leftarrow \{\}$ \; \texttt{// Store phase tables}

    \For{$x \leftarrow 0$ \KwTo $\lceil \frac{\text{len}(\underline{e})}{w_e} \rceil - 1$}{
        $\underline{e}_{\text{window}} \leftarrow \underline{e}$.\POP(size=$w_e$) \;

        \For{$y \leftarrow 0$ \KwTo $\lceil \frac{\text{len}(\underline{m})}{w_m} \rceil - 1$}{
            $\underline{m}_{\text{window}} \leftarrow \underline{m}$.\POP(size=$w_m$) \;
            table $\leftarrow$ \LOOKUPTABLE($g$, $x$, $w_e$, $y$, $w_m$) \;

            \QRAMLOOKUP{$\underline{m}_{\text{window}} || \underline{e}_{\text{window}}, \text{table}, \underline{l}$} \;
            \INPLACEADDMOD{$\underline{t}, \underline{l}$} \;

            $F[y] \leftarrow$ \CPUT{$w_e, w_m, \text{table}, \underline{l}$} \;
        }

        \INITUNARY{$\underline{e}_{\text{window}}, \underline{l}[0:2^{w_e}]$} \;

        \For{$y \leftarrow 0$ \KwTo $\lceil \frac{\text{len}(\underline{m})}{w_m} \rceil - 1$}{
            $\underline{m}_{\text{window}} \leftarrow \underline{m}$.\POP(size=$w_m$) \;
            $\mathcal{H}(\underline{l}[0:2^{w_e}])$ \;
            \LOOKUP{$\underline{m}_{\text{window}}, F[y], \underline{l}[0:2^{w_e}]$} \;
            $\mathcal{H}(\underline{l}[0:2^{w_e}])$ \;
        }

        \texttt{reverse INIT\_UNARY($\underline{e}_{\text{window}}, \underline{l}[0:2^{w_e}]$)}
        
        \tcp{Swap target and multiplication registers}
            $\underline{t}, \underline{m} \leftarrow \underline{m}, \underline{t}$\;
            
        $\;\;\;\vdots$
        
        $\;\;\;\vdots$ uncompute register $m$
        
        $\;\;\;\vdots$
    }
    
}
\end{algorithm}
  \caption{Deferred uncomputation}
  \label{lst:windowed_quantum_modular_exponentiation_du}
\end{table}

\end{document}